\documentclass[aps,prb,reprint,superscriptaddress]{revtex4-1}
\usepackage{braket}
\usepackage{graphicx}
\usepackage{amsmath}
\usepackage{amssymb}
\usepackage{bbold}
\usepackage{bm}
\usepackage[usenames]{color}

\renewcommand{\vec}[1]{\mathbf{#1}}

\begin{document}


\title{Direct coupling between charge current and spin polarization by extrinsic mechanisms in graphene}

\author{Chunli Huang}
\affiliation{
 Division of Physics and Applied Physics, School of Physical and Mathematical Sciences,
Nanyang Technological University, Singapore 637371, Singapore
}
\affiliation{Department of Physics, National Tsing Hua University, Hsinchu 30013, Taiwan}

\author{Y. D. Chong}
\affiliation{
 Division of Physics and Applied Physics, School of Physical and Mathematical Sciences,
Nanyang Technological University, Singapore 637371, Singapore
}

\author{Miguel A. Cazalilla}
\affiliation{Department of Physics, National Tsing Hua University and National Center for Theoretical Sciences (NCTS), Hsinchu 30013, Taiwan}

\date{\today}

\begin{abstract}
Spintronics---the all-electrical control of the electron spin for quantum or classical information storage and processing---is one of the most promising applications of the two-dimensional material graphene.  Although pristine graphene has negligible spin-orbit coupling (SOC), both theory and experiment suggest that SOC in graphene can be enhanced by extrinsic means, such as functionalization by adatom impurities.  We present a theory of transport in graphene that accounts for the spin-coherent dynamics of the carriers, including hitherto-neglected spin precession processes taking place during resonant scattering in the dilute impurity limit. We uncover a novel ``anisotropic spin precession'' (ASP) scattering process in graphene, which contributes a large current-induced spin polarization and modifies the standard spin Hall effect. ASP scattering arises from two dimensionality and extrinsic SOC, and apart from graphene, it can be present in other 2D materials or in the surface states of 3D materials with a fluctuating SOC. Our theory also yields a comprehensive description of the spin relaxation mechanisms  present in adatom-decorated graphene, including Elliot-Yafet and D'yakonov-Perel relaxation rates, the latter of which can become an amplification process in a certain parameter regime of the SOC disorder potential. Our work provides theoretical foundations for designing future graphene-based integrated spintronic devices. 
\end{abstract} 

\maketitle

\section{Introduction}
The role of quantum spin coherence in graphene spin transport is presently poorly understood. It has been neglected in previous theoretical treatments based on semiclassical transport equations~\cite{ferreira2014extrinsic,hy2015extrinsic}.  Regarding the magnitude of the spin relaxation time, 
which determines the degree of quantum coherence, 
there is substantial disagreement between theory and experiment \cite{Pesin12,Ochoa12,FertMRS14}, as well as between different experiments~\cite{Tombros2007,electron2011wu,zomer2012long,dlubak2012highly}. Nevertheless, it seems that graphene exhibits fairly long spin relaxation times compared to metals, making it a promising material for \emph{passive} spintronics, i.e.~long-distance transport of spin currents~\cite{Tombros2007,electron2011wu,Pesin12,zomer2012long,Ochoa12,dlubak2012highly,stephan2014,FertMRS14}. 

There is also a solid body of theoretical~\cite{netoguinea09,weeks2011engineering, pachoud2014scattering,fabian2015spin} and experimental \cite{marchenko2012giant,calleja2015natphy,balakrishnan_colossal,balakrishnan2014giant} work indicating that graphene can exhibit strong extrinsic SOC induced by proximity to adatom impurities or metallic substrates. This suggests that graphene-based devices can also play an important role in \emph{active} spintronics.  Extrinsic SOC induced by proximity to metals and metal clusters has been detected experimentally in graphene, via angle-resolved photoemission~\cite{marchenko2012giant}, scanning tunneling spectroscopy~\cite{calleja2015natphy}, and spin transport~\cite{balakrishnan_colossal,balakrishnan2014giant}.  In particular, Balakrishan \emph{et al}.~\cite{balakrishnan2014giant} have reported observing the spin Hall effect (SHE) in graphene devices sparsely decorated with copper clusters (residues found in graphene grown by chemical vapor deposition); they found that the spin Hall angle (the ratio of spin current to charge current) was $\theta_{\mathrm{sH}} \sim 0.1$, comparable to transition metals~\cite{Kimura2007,sinova2015spin} and 2D transition metal dichalcogenides~\cite{Wang2012,Qian2014}. Although other groups have reproduced the nonlocal resistance measurements in adatom-decorated graphene, they have failed to observe the expected modulation with in-plane magnetic field (Hanle precession)~\cite{neutral2015wang,kaverzin2015electron}.
Other recent experiments \cite{oshima2014observation,mendes2015spin,dushenko2016gate} have demonstrated spin-charge conversion in graphene by spin-pumping, and some~\cite{oshima2014observation,dushenko2016gate} reported values of $\theta_{\mathrm{sH}}$ many orders of magnitude below what was obtained in earlier Hall-bar devices \cite{balakrishnan2014giant}.  In our view, the confusing experimental situation calls for a more detailed theoretical analysis of how spin currents and charge currents are coupled by extrinsic SOC in graphene.
 
A semi-classical theory of the SHE in graphene with extrinsic SOC has been developed by Ferreira \textit{et al.}~\cite{ferreira2014extrinsic}  That work showed that the spin Hall angle can be enhanced in graphene by resonant skew scattering, and that the enhancement is much stronger than in bulk metals~\cite{Fert_resonant1} due to graphene's 2D Dirac density of states~\cite{wehling2009adsorbates,katsnelson2012graphene}.  However, the theory neglected quantum spin coherence, describing the carriers using distinct ``spin-up'' and ``spin-down'' distributions~\cite{ferreira2014extrinsic}. As shown below, this can be strictly  justified only when the extrinsic SOC is purely of the spin-conserving (Maclure-Yafet-Kane-Mele) type~\cite{KaneMeleQSHE1995}.  Rashba-type SOC, which favors spin polarization parallel to the graphene plane (and appears when reflection symmetry about the plane is broken), gives rise to coherent spin precession and spin relaxation processes that have not been accounted for, which could lead to qualitative devitations from the results of Ref.~\onlinecite{ferreira2014extrinsic}.  Indeed, numerous tight-binding and density functional theory studies have found that adatom impurities induce both types of SOC, with comparable strengths \cite{netoguinea09,weeks2011engineering,stabilizing2012hua,fabian2013spin,fabian2015spin}.

\begin{figure*}[ht]
\includegraphics[width=0.95\textwidth]{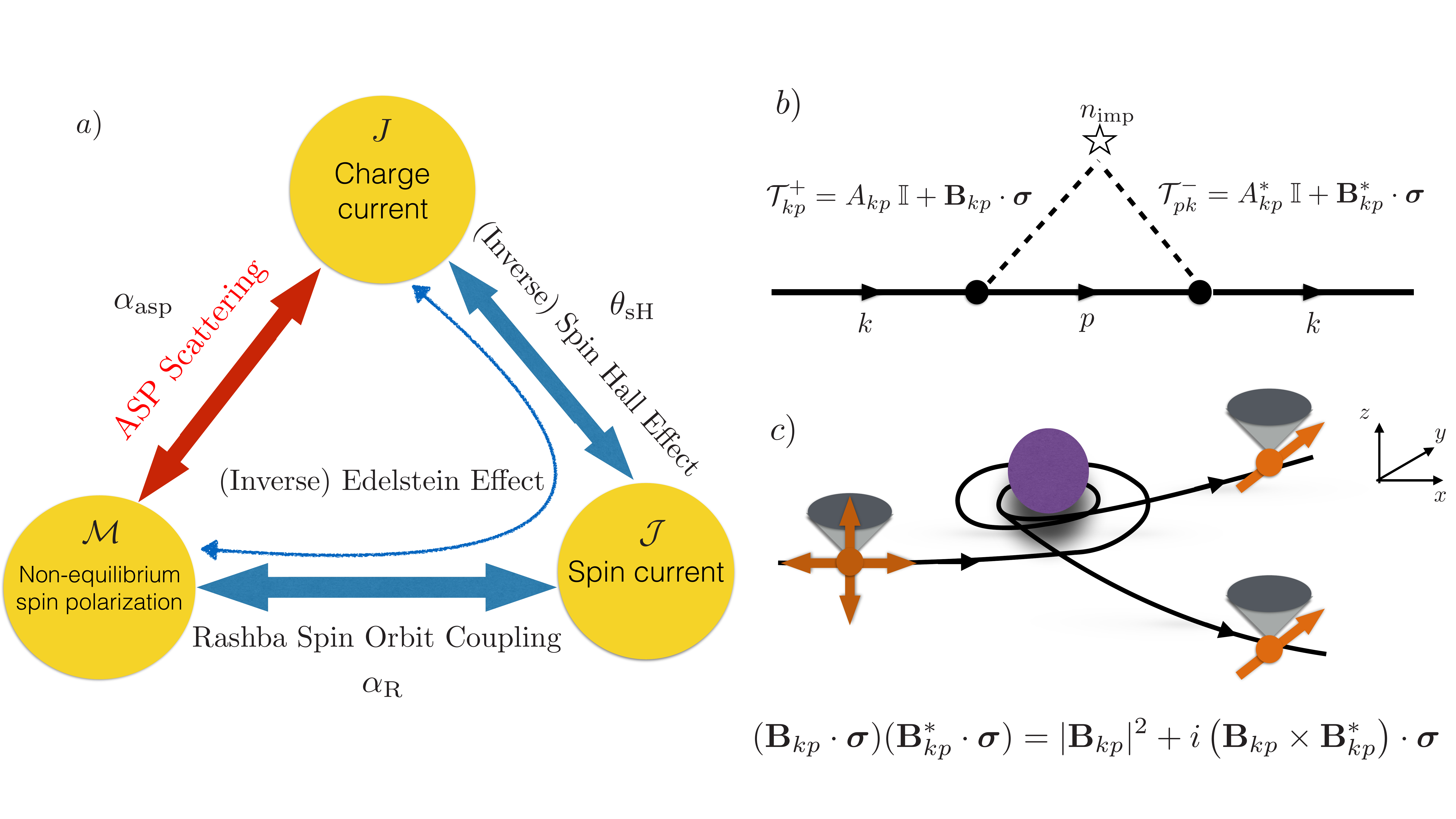}
 \caption{\label{fig:triangle}
(a) Relationship between the three basic macroscopic transport quantities (charge current $J$, spin current $\mathcal{J}$, and magnetization $\mathcal{M}$) in systems with time-reversal symmetric microscopic dynamics. The coupling between $J$ and $\mathcal{J}$ is governed by the spin Hall angle $\theta_{\mathrm{sH}}$, and the coupling between $\mathcal{J}$ and $\mathcal{M}$ is governed by the Rashba scattering rate $\alpha_{\mathrm{R}}$; together, these yield the Edelstein effect. The anisotropic spin precession (ASP) scattering rate, $\alpha_{\mathrm{asp}}$, describes a novel direct coupling between $J$ and $\mathcal{M}$.  (b) Feynman diagram corresponding to scattering events linear in the impurity density $n_{\text{imp}}$. The scattering amplitude involves the product of $T$-matrix elements, and different terms in the product describe ASP scattering, skew scattering, Rashba scattering, Drude relaxation, and Elliott-Yafet spin relaxation. (c) Schematic of ASP scattering, which arises from the terms in the product given by $\boldsymbol{c}_{3}= i \boldsymbol{B}_{kp} \times \boldsymbol{B}_{kp}^{*}$. For example, the product of $(\boldsymbol{B}_{kp})^z\sigma^z$ and $(\boldsymbol{B}^*_{kp})^x\sigma^x$ can cause a randomly polarized spin to align in the $y$ direction. The former skews while the latter flips the electron spin.  This gives rise to spin alignment in the $y$ direction, $\mathcal{M}^y \ne 0$, but no net transverse spin current $\mathcal{J}_y^z$. 
}
\end{figure*}

Numerical quantum transport simulations are an alternative approach to studying extrinsic SOC in graphene, and are capable of accounting for quantum spin coherence.  So far, however, simulations based on the Kubo formalism \cite{rappoport,stephan2016revisit} have been limited to impurity densities of $\gtrsim 10\%$, significantly higher than in typical experiments~\cite{balakrishnan_colossal,balakrishnan2014giant,kaverzin2015electron} ($\lesssim 1\%$).  It is uncertain whether these numerical methods can be extrapolated to the dilute impurity regime, due to the different, density dependent mechanisms involved in the SHE~\cite{sinova2015spin,mirco2016phase,mirco2016quantum}.

This paper presents an analytical theory of spin-coherent transport for extrinsic SOC in graphene, based on the linearized Quantum Boltzmann Equation (QBE).  It incorporates coherent spin dynamics into the transport equations by treating both types of extrinsic SOC on equal footing, and is directly applicable to the experimentally-relevant limit of strongly-scattering but dilute impurities \cite{balakrishnan_colossal,balakrishnan2014giant,kaverzin2015electron}.  From the theory, we uncover a novel extrinsic SOC scattering process, ``anisotropic spin precession'' (ASP) scattering, which involves a combination of skew and spin-flip scattering.  ASP scattering directly couples non-equilibrium spin polarization and charge current [cf.~Fig.~\ref{fig:triangle}(a)], and is distinct from other previously-studied scattering processes that couple non-equilibrium spin polarization $\mathcal{M}$, charge current $J$, and spin current $\mathcal{J}$.

One of the most striking predictions of the theory is that graphene can exhibit a sizable current-induced spin polarization (CISP).  CISP, also known as the inverse spin-galvanic 
effect \cite{cisp2004kato,sih2005spatial,cisp2014dydra}, refers to the production of non-equilibrium spin polarization (i.e.~magnetization) by passing a charge current through a material.  So far, the mechanism that has been identified as the cause of CISP is the Edelstein effect \cite{Edelstein1990233,pikus1991spin,shen2014theory}, in which a charge current $J$ is first converted into a spin current $\mathcal{J}$ via the SHE, and $\mathcal{J}$ is then converted into spin polarization, $\mathcal{M}$, by Rashba SOC~\cite{raimondi2012su2}.  However, as we show below, in graphene doped with SOC impurities, CISP arises from both an extrinsic version of the Edelstein effect and ASP scattering; and the latter is dominant when the Rashba SOC induced by the impurities is strong.

Apart from CISP, ASP scattering also contributes to the spin current $\mathcal{J}$, and the size of its contribution can be comparable to the standard SHE contribution caused by skew scattering.   In particular, the ASP scattering contribution is distinct from side-jump scattering~\cite{mirco2016phase,mirco2016quantum}, which is another mechanism that contributes to the SHE in graphene.  ASP scattering also gives important corrections to spin relaxation processes, and particularly the D'yakonov-Perel (DP) relaxation time.

The rest of the article is organized as follows. Sec.~\ref{sec:results} summarizes our most important results, including the linear response equation relating charge current, spin current, and non-equilibrium spin polarization to the applied electric field.  We also discuss some of the experimental implications from our theory and clarify the spin relaxation mechanisms in graphene with SOC disorder. Sec.~\ref{sec:QBE} provides a brief summary of the QBE, emphasizing the structure of the collision integral.  Sec.~\ref{sec:model} explains how the linearized QBE can be solved after introducing a microscopic model for the SOC disorder potential and an \emph{ansatz} for the electron distribution function.  Finally, in Sec.~\ref{sec:summary} we close the article with a summary and outlook. Key technical details and the detailed derivation of the QBE from the equation of motion for the density matrix are given in the Appendix.
 
\section{Results} \label{sec:results}

In this section, we present the theory's main results, leaving a discussion of its derivation to Sec.~\ref{sec:QBE}.  When both types of extrinsic SOC (Maclure-Yafet-Kane-Male and Rashba) are present in adatom decorated graphene, and quantum spin coherence is accounted for, the linear response of the system becomes qualitatively different from the semi-classical descriptions previously developed in Refs.~\cite{ferreira2014extrinsic,hy2015extrinsic}.  Specifically, let us consider an electric field $E_x$ applied along the $\hat{x}$ direction,  the graphene plane being the $\hat{x}$-$\hat{y}$ plane.  The response of the system in terms of the longitudinal charge current $J_x$, transverse spin current $\mathcal{J}_y^z$, and magnetization $\mathcal{M}^y$ (in rescaled units; see Sec.~\ref{sec:QBE}) takes the form:
\begin{multline} \label{eq:transport}
\begin{pmatrix}
J_x \\ \mathcal{J}_y^z \\ \mathcal{M}^y \end{pmatrix}
=
\begin{pmatrix}
0 & \theta_{\mathrm{sH}} &  \tau_\mathrm{D} \alpha_{\mathrm{asp}} \\
-\theta_{\mathrm{sH}} & 0 &  \tau_\mathrm{D} \alpha_{\mathrm{R}} \\
\tau_{\mathrm{EY}}  \alpha_{\mathrm{asp}} 
& - \tau_{\mathrm{EY}} \alpha_{\mathrm{R}} & 0 
\end{pmatrix}
\begin{pmatrix}
J_x \\
\mathcal{J}_y^z \\
\mathcal{M}^y
\end{pmatrix}
\\+ \sigma_{\mathrm{D}}
\begin{pmatrix} E_x \\ 0 \\ 0 \end{pmatrix}.
\end{multline}
\noindent
The first term on the right describes the coupling of spin and charge, and the second term describes the out-of-equilibrium drive ($\sigma_\mathrm{D}$ is the Drude conductivity).  In the (dimensionless) coupling matrix, $\tau_{\text{D}}$ and $\tau_{\text{EY}}$ are the Drude relaxation time and Elliott-Yafet spin relaxation time (see Sec.~\ref{sec:Spin relaxation mechanisms}); $\alpha_{\mathrm{R}}$ is the scattering rate induced by Rashba SOC; and $\theta_{\text{sH}} $ is the spin Hall angle. In the experimentally relevant dilute-impurity regime \cite{balakrishnan2014giant, neutral2015wang}, the dominant contribution to the SHE arises from skew scattering \cite{ferreira2014extrinsic,sinova2015spin}, and thus $\theta_{\text{sH}}=\tau_{\mathrm{D}}\alpha_{\mathrm{sk}}$ where $\alpha_{\mathrm{sk}}$ is the skew scattering rate. These relationships are depicted in Fig.~\ref{fig:triangle}(a).

The matrix elements that directly couple $J_x$ with $\mathcal{M}^y$ are a novel outcome of the theory. They are governed by the ASP scattering rate, $\alpha_{\text{asp}}$. As shown in Fig.~\ref{fig:triangle}(c), ASP scattering arises from quantum interference between skew scattering and spin flip scattering: in a single scattering event, the electron is skewed and then flipped (or vice versa), and this results in a net spin polarization in the plane.  As discussed in Sec.~\ref{sec:QBE}, the existence of ASP scattering is fundamentally due to the existence of a special axis (the out-of-plane direction) in 2D materials, which breaks rotational symmetry for rotations about this axis. ASP scattering vanishes in 3D materials possessing time-reversal and 3D rotational symmetries (such as the system studied by Lifshits and Dyakonov in Ref.~\onlinecite{dyakonov_swapcurrent2009}). As shown in Appendix \ref{app:sym}, ASP scattering can also occur in 2D electron gases in quantum wells, and in other spin-orbit coupled electron systems when the full 3D rotational symmetry is broken, such as interfaces.  

Note that the coupling between $J$ and $\mathcal{J}$ (the SHE) and between $\mathcal{J}$ and $\mathcal{M}$ are both odd under Onsager reciprocity. In other words, in the matrix in Eq.~\eqref{eq:transport}, the $(1,2)$ and $(2,1)$ elements have opposite signs, and likewise the $(2,3)$ and $(3,2)$ elements have opposite signs.  This is consistent with the fact that $J$ and $\mathcal{M}$ are odd, and $\mathcal{J}$ is even, under time-reversal. On the other hand, the new ASP scattering induced couplings between $J$ and $\mathcal{M}$ are even under Onsager reciprocity; correspondingly, the $(1,3)$ and $(3,1)$ matrix elements have the same sign. This has important consequences for the ASP contributions to spin relaxation, as discussed in Section \ref{sec:Spin relaxation mechanisms}.

\subsection{Current-induced spin polarization}\label{sec:cisp}

\begin{figure}
  \includegraphics[width=8cm]{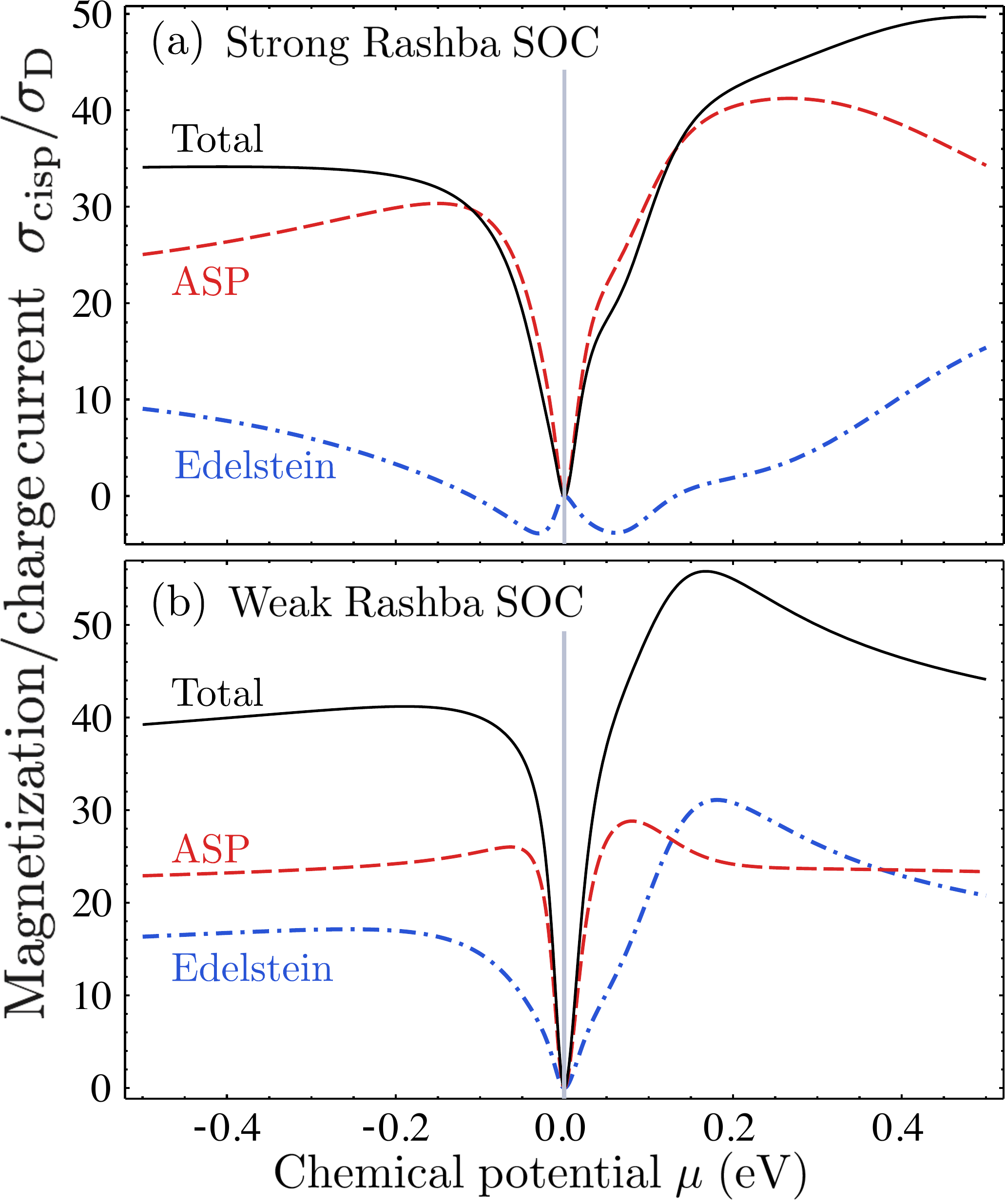}
  \caption{\label{fig:sigmaee}
  	Ratio of magnetization to charge current at zero temperature, plotted versus the chemical potential $\mu$ (measured from the Dirac point). The magnetization is measured in Bohr magnetons.  The top panel is plotted for strong Rashba SOC with bare Rashba potential strength of $30$~meV while the bottom panel is for weak Rashba SOC with bare Rashba potential strength of $10$~meV . In both cases, the strength of the spin-conserving and scalar potentials are set to $10$~meV and $80$~meV, respectively, in line with previous theoretical studies.~\cite{ferreira2014extrinsic,balakrishnan2014giant}  See appendix~\ref{app:micro_model} for details.}
\end{figure}

To compute the CISP, we expand the last row in Eq.~\eqref{eq:transport} to first order in $\alpha_{\mathrm{R}}\theta_{\mathrm{sH}}$ and $\alpha_{\mathrm{asp}}$.  This gives $\mathcal{M}^{y}=\sigma_{\mathrm{cisp}}E_{x}$, where
\begin{equation}\label{eq:edelstein}
\sigma_{\mathrm{cisp}}=\sigma_{\mathrm{D}} \left(\theta_{\mathrm{sH}}\alpha_{\mathrm{R}}  +  \alpha_{\mathrm{asp}} \right) \tau_{s}.
\end{equation}
Here $\tau_{s}^{-1}= \tau_{\text{EY}}^{-1}+\tau_{\text{DP}}^{-1}$ is the total spin relaxation time (see Section~\ref{sec:Spin relaxation mechanisms}).  Eq.~(\ref{eq:edelstein}) shows that there are two distinct mechanisms contributing to the CISP: (i) the \emph{extrinsic} Edelstein effect which is a two-step process associated with $\theta_{\mathrm{sH}}\alpha_{\mathrm{R}}$, and (ii) the ASP scattering which is associated with $\alpha_{\mathrm{asp}}$.  The first term is formally identical to the Edelstein effect found in the 2D electron gas, \cite{shen2014theory} with one important difference: the effect here is of extrinsic origin. Unlike the \emph{intrinsic} Eldestein effect \cite{Edelstein1990233,pikus1991spin,shen2014theory}, which arises from a spatially uniform Rashba SOC, the \emph{extrinsic} Eldestein effect arises from impurity scattering.  Its strength is determined by the Rashba scattering rate $\alpha_{\mathrm{R}}$, which depends on the chemical potential $\mu$. 

The second term in Eq.~\eqref{eq:edelstein} describes the enhancement of the magnetization by ASP scattering. Normally, since ASP scattering is present even in the first Born approximation, we would expect it to dominate over the Edelstein effect, which appears only in the third Born approximation. However, in the model calculations below, all the scattering rates are computed to all orders in the impurity potential strength.  This results in the two contributions being comparable in magnitude for most $\mu$.

In Fig.~\ref{fig:sigmaee}, we plot the dimensionless ratio $\sigma_{\text{cisp}}/\sigma_{\mathrm{D}}$, which serves as a figure of merit for CISP, against the chemical potential $\mu$. It exhibits a peak for positive $\mu$ due to resonant spin-coherent scattering ($\sim 100$\,meV based on our choice of SOC impurity potential; see Sec.~\ref{sec:model}).  ASP scattering gives the dominant contribution to CISP when the impurity Rashba SOC is large.

Both the extrinsic Edelstein and ASP scattering contributions to CISP are proportional to the total spin relaxation time. Due to the specific features of graphene, this implies that the net magnetization, normalized by the charge current, can be enhanced both by the resonant enhancement of the SHE~\cite{ferreira2014extrinsic}, and the long spin relaxation times characteristic of graphene~\cite{zomer2012long,dlubak2012highly}.

\subsection{Current-induced spin current}

\begin{figure}
  \includegraphics[width=8cm]{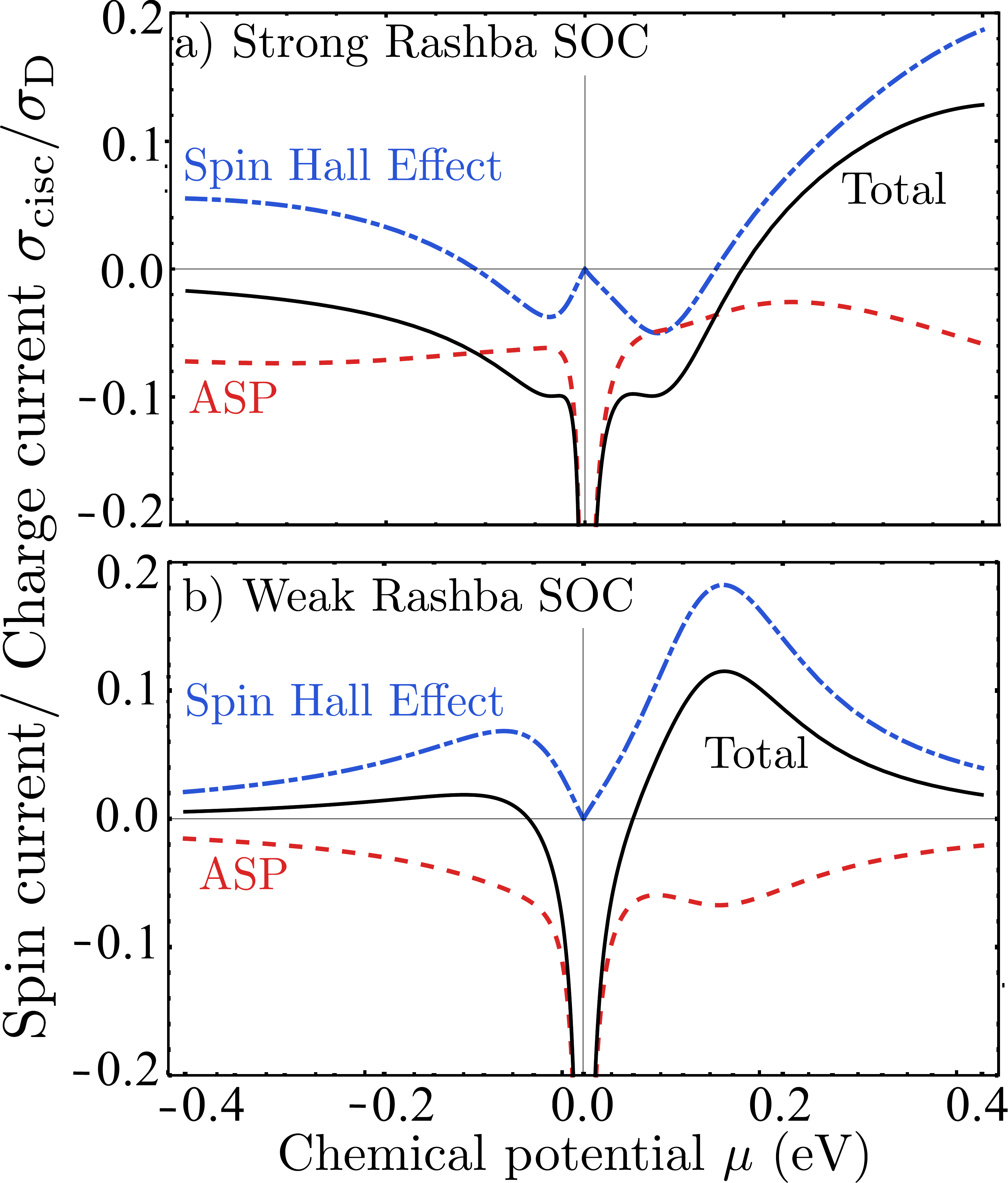}
  \caption{\label{fig:cisc}
  	Ratio of spin current to charge current at zero temperature, plotted versus the chemical potential $\mu$ (measured from the Dirac point). In the close vicinity of Dirac point, the contribution of ASP scattering goes to number much larger than one. This is because $\tau_{\mathrm{EY}}$ goes to infinity much faster than $\alpha_{\mathrm{asp}}$ goes to zero, an artefact of the theory. The parameters used here are the same as in Fig.~\ref{fig:sigmaee}}
\end{figure}

In addition to the magnetization, we can compute the spin current from the second row of Eq.~\eqref{eq:transport}.  To first order in $\theta_{\mathrm{sH}}$ and $\alpha_{\mathrm{asp}}\alpha_{\mathrm{R}}$, we find that $\mathcal{J}^{z}_{y}=\sigma_{\mathrm{cisc}}E_{x}$, where
\begin{equation}\label{eq:SHE}
\sigma_{\mathrm{cisc}}= \sigma_{\mathrm{D}}\Big(-\theta_{\mathrm{sH}}+  \left( \alpha_{\mathrm{R}}\tau_{\mathrm{D}}  \right) \left( \alpha_{\mathrm{asp}}\tau_{\mathrm{EY}}  \right)\Big).
\end{equation}
The first term in Eq.~\eqref{eq:SHE} is the conventional spin Hall conductivity arising from skew scattering.  The second term arises from a combination of ASP scattering and Rashba scattering. If Rasba SOC is absent, the ``skewness ratio'' $\sigma_{\mathrm{cisc}}/ \sigma_{\mathrm{D}}$ reduces to $\theta_{\mathrm{sH}}$, conventionally known as the spin Hall angle.  Note that both terms are independent of the impurity density, unlike the quantum side-jump contribution to the spin Hall current \cite{mirco2016phase,mirco2016quantum}.  (Side-jump is not included in the present theory.)

Fig.~\ref{fig:cisc} shows $\sigma_{\mathrm{cisc}}/ \sigma_{\mathrm{D}}$ versus $\mu$.  For strong impurity-induced Rashba SOC, the skewness ratio is enhanced for $\mu$ near the scattering resonance ($\sim 100$\,meV; see Sec.~\ref{sec:model}), with the skew scattering and ASP/Rashba contributions having the same sign.  However, for weak impurity-induced Rashba,  the two contributions can have opposite signs, diminishing the total spin current response.  This is consistent with a previous semiclassical (non-spin-coherent) calculation which found a similar reduction under weak Rashba SOC disorder \cite{ferreira2014extrinsic}.  These plots also indicate that at small $\mu$, there is a sharp increase in the skewness ratio coming from the ASP/Rashba scattering contribution.  Specifically, as $\mu$ tends to zero, $\alpha_{\mathrm{asp}}$ vanishes more slowly than the Elliot-Yafet spin relaxation time $\tau_{\mathrm{EY}}$ increases. However, at very small values of $\mu$ ($\sim 10$\,meV), our theory becomes unreliable due to multiple impurity scattering and interband coherence effects (see Sec.~\ref{sec:model}). 
 It is interesting to note that the Edelstein contribution to CISP (blue dotted line in Fig.~\ref{fig:sigmaee}) tracks the SHE contribution to CISC (blue dash-dotted line in Fig.~\ref{fig:sigmaee}. This is   because the Edelstein effect is a two-step conversion process, see Fig.~\ref{fig:triangle}a.

The new ASP scattering contribution to the spin current calls for a need to revise the existing SHE theory \cite{abanin2009nonlocal} which has been customarily employed to fit the nonlocal resistance data in Hall bar spin-transport experiments~\cite{balakrishnan2014giant,balakrishnan_colossal,neutral2015wang,kaverzin2015electron}. This is especially true when the impurity-induced Rashba is large, as it may be the case of hydrogenated graphene \cite{balakrishnan2014giant,kaverzin2015electron}; such an analysis will be presented elsewhere~\cite{in-prep}.  The present theory may also have important implications for understanding recent spin pumping experiments in CVD graphene involving the inverse of CISP \cite{mendes2015spin} and the inverse of current-induced spin current~\cite{oshima2014observation,dushenko2016gate}.

\subsection{Spin relaxation}
\label{sec:Spin relaxation mechanisms}

\begin{figure*}
  \includegraphics[width=\textwidth]{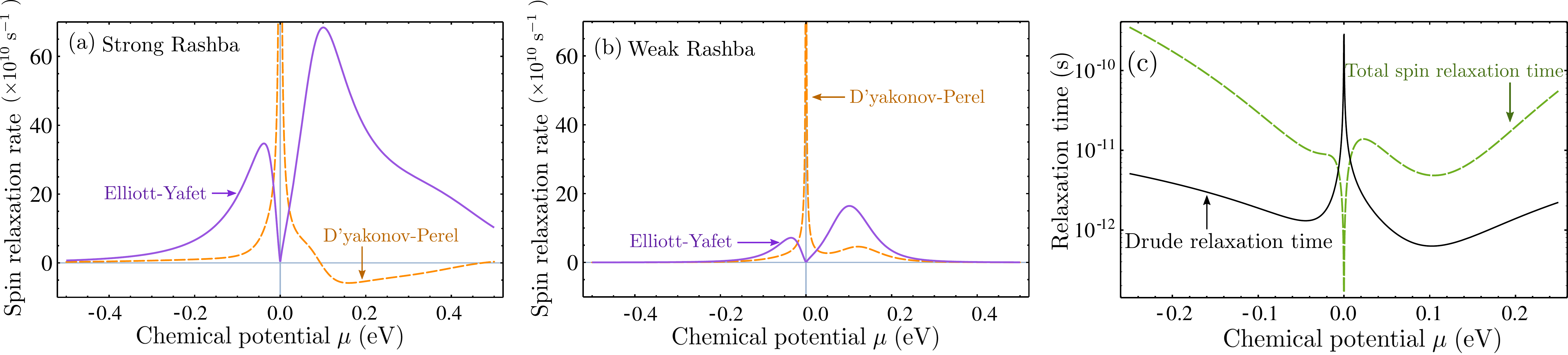}
  \caption{\label{fig:eydp} (a)--(b) Spin relaxation rates of the Elliott-Yafet type, $\tau_{\mathrm{EY}}^{-1}$, and Dyakonov-Perel type, $\tau_{\mathrm{DP}}^{-1}$, for different Rashba SOC strengths. Note that the DP spin relaxation rate becomes negative for a window of chemical potential where $\alpha_{\mathrm{R}}^2 <\alpha_{\mathrm{sk}}^2$. Both scattering rates show resonant enhancements near the vicinity of Dirac point (i.e.~$\mu=0$). At large $\mu$, the EY scattering rate dominates.  (c) Relaxation time versus $\mu$ for weak Rashba SOC.    In these calculations, we assume identical impurities and zero temperature; for a random distribution of impurity strengths and/or finite temperatures, these resonant features will be further smoothed out. The parameters are the same as in Fig.~\ref{fig:sigmaee}}
\end{figure*}
 
The theory derived in Sec.~\ref{sec:QBE} also allows us to obtain the spin relaxation rate arising from SOC disorder, $\tau^{-1}_{s}$. In the stationary state, this relaxation rate is obtained by solving Eq.~\eqref{eq:transport} for the spin polarization $\mathcal{M}^y$, as described in Section~\ref{sec:cisp}. We find that $\tau_s^{-1}$ receives two contributions, which add up by Matthiessen's rule
\begin{equation}
\tau^{-1}_s = \tau^{-1}_{\mathrm{EY}} + \tau^{-1}_{\mathrm{DP}}.
\end{equation}
The two contributions can be identified as Elliott-Yafet (EY) relaxation \cite{Ochoa12} and D'yakonov-Perel (DP) relaxation~\cite{vzutic2004spintronics}.  The rates are derived to be
\begin{align}
\frac{1}{\tau_{\text{EY}} } =&  \frac{C}{\tau_{\mathrm{D}} },
\label{tauey} \\ 
\frac{1}{\tau_{\text{DP}}} =&  \tau_{\mathrm{D}}(\alpha_{\mathrm{R}}^2-\alpha_{\mathrm{asp}}^2), \label{taudp} 
\end{align}
\noindent
where $\tau_{\mathrm{D}}$ is the Drude relaxation (elastic scattering) time and $C > 0$ depends on the microscopic model (see Sec.~\ref{sec:model}). Since our theory assumes that the SOC disorder stems from localized impurities (e.g.~adatoms \cite{balakrishnan_colossal,balakrishnan2014giant, kaverzin2015electron}), the EY relaxation is caused by spin-flip scattering events.  This form of   EY spin relaxation is akin to the one considered by Lifshits and Dyakonov in Ref.~\onlinecite{dyakonov_swapcurrent2009}, but different from other models (e.g.~Ref.~\onlinecite{Ochoa12}) where the Rashba SOC is assumed to be uniform.

The DP relaxation time, $\tau_{\mathrm{DP}}$, is related to spin precession and was previously understood to arise from Rashba SOC scattering. We argue that this understanding is incomplete: $\tau_{\mathrm{DP}}$ also receives a contribution from ASP scattering.  It can be seen from Eq.~\eqref{taudp} that, unlike the EY relaxation time $\tau_{\text{EY}}$ which is strictly positive, the sign of the DP relaxation time $\tau_{\text{DP}}$ depends on the competition between $\alpha_{\mathrm{R}}$ and $\alpha_{\mathrm{asp}}$.  As noted above, ASP scattering is an Onsager even process. As a result, its contribution to $\tau_{\text{DP}}$  is  negative. When $\alpha_{\mathrm{R}}^2 > \alpha_{\mathrm{asp}}^2$, $\tau_{\text{DP}}$ describes a spin relaxation process, whereas when $\alpha_{\mathrm{R}}^2 < \alpha_{\mathrm{asp}}^2$, it describes spin amplification (the total spin relaxation time, however, remains strictly positive, cf.~Fig.~\ref{fig:eydp}). This shows that Rashba SOC can either randomize or align the spin polarization, depending on the microscopic details of the system. We also find that the total spin relaxation time is minimum near zero-doping, which agrees well with the trend observed in the experiments~\cite{spinrelaxation2011han,linear2009jozsa}, and suggesting that resonant scattering with SOC disorder is an important source of spin relaxation at low temperatures. 

In Fig.~\ref{fig:eydp}, the EY and DP spin relaxation rates are plotted against the chemical potential $\mu$, for strong and weak Rashba SOC. We find that EY relaxation is the dominant mechanism for spin relaxation at large $\mu$, but DP spin relaxation becomes important at small $\mu$ since it is inversely proportional to the Drude relaxation time. Fig.~\ref{fig:eydp}(c) shows that the total spin relaxation time $\tau_{s}$ approaches a minimum as the Drude relaxation (elastic scattering) time $\tau_{\mathrm{D}}$ peaks near $\mu = 0$ (the Dirac point). This agrees with the experimental observations of spin relaxation in exfoliated graphene~\cite{spinrelaxation2011han} and CVD graphene,~\cite{kamalakar2015long} which indicate that both EY and DP spin relaxation mechanisms are present in graphene.

Finally, we note that Eqs.~\eqref{eq:transport}--\eqref{taudp} follow from the general form of the QBE within the linear response regime, and are independent of the underlying microscopic scattering model.  The parameters $\{\tau_{\mathrm{D}}, \tau_{\mathrm{EY}}, \alpha_{\mathrm{sk}}, \alpha_{\mathrm{R}}, \alpha_{\mathrm{asp}}\}$ entering into these equations all depend on the chemical potential, the impurity density, and the SOC impurity potential.  Their actual values must be derived from a microscopic scattering model, which in turn is fitted to \textit{ab initio} calculations and/or experimental measurements.  Details of this derivation are given in Sec.~\ref{sec:model} and the Appendix.

\section{Quantum Boltzmann Equation in the strong and dilute disorder regime} \label{sec:QBE}

In this section, we discuss the quantum transport equation that leads to the linear response matrix equation \eqref{eq:transport}.  To capture the coherent quantum dynamics of electron spins in disordered graphene, we use the method of Kohn and Luttinger~\cite{KohnLuttingerBTE} to derive a quantum kinetic equation.  The collision integral derived is first-order in the impurity concentration $n_{\mathrm{imp}}$, but exact to all orders in the strength of the single-impurity potential.  Details of this formalism are given in the Appendix \ref{sec:app_QBE}.  Unlike the original Kohn-Luttinger treatment, we keep track of the quantum spin coherence by using a $2 \times 2$ density matrix distribution $n_k(t)$.  The deviation from equilibrium distribution is given by $\delta n_{k}(t) = n_k(t) - n^0_k$, where $n^0_k = f_{\mathrm{FD}}(\epsilon_k)\: \mathbb{1}$ is the equilibrium distribution, $\mathbb{1}$ is the $2\times 2$ unit matrix, $\epsilon_k =  v_F k$   ($\epsilon_k = - v_F k$) is the dispersion relation of electrons (holes) in graphene, and $f_{\mathrm{FD}}(\epsilon) = (e^{(\epsilon-\mu)/k_B T}+1)^{-1}$ is the Fermi-Dirac distribution at absolute temperature $T$ and chemical potential $\mu$.  This results in a linearized quantum Boltzmann equation (QBE) which describes how $\delta n_{k}$ reacts to applied electric and magnetic fields, $\boldsymbol{E}(t)$ and $\boldsymbol{\mathcal{H}}(t)$:
\begin{align}
\label{eq:QBE}
\partial_t \delta n_k + \frac{i}{\hbar} \gamma \left[ \delta n_k , \boldsymbol{s}\cdot \boldsymbol{\mathcal{H}}(t)\right] 
+ e \boldsymbol{E}(t)\cdot \frac{\boldsymbol{\nabla}_{k} n^0_k}{\hbar} =
\mathcal{I}\left[ \delta n_k \right].
\end{align}
\noindent
Here, $\boldsymbol{s} = \frac{\hbar}{2} \boldsymbol{\sigma}$ is the electron spin operator ($\boldsymbol{\sigma} = (\sigma^x, \sigma^y, \sigma^z)$ are the Pauli matrices) and $\gamma$ is the gyromagnetic ratio. In deriving Eq.~\eqref{eq:QBE}, we have assumed that the external (electric and magnetic) fields vary slowly in time compared to the time scale of $\hbar/\mu$.

To leading order in $n_{\mathrm{imp}}$, the collision integral is
\begin{align}
\label{eq:coll-int}
\mathcal{I}[\delta n_{k}] = \frac{ i}{\hbar}  [\delta n_{k} ,\Sigma^R_{k} ] +\frac{2\pi n_{\mathrm{imp}}}{\hbar} \sum_{\boldsymbol{p}} \delta(\epsilon_{k}-\epsilon_{p})  \nonumber \\ 
 \times \left( \mathcal{T}^{+}_{kp} \delta n_{p} \mathcal{T}^{-}_{pk}-
\frac{\mathcal{T}^{+}_{kp}\mathcal{T}^{-}_{pk}\delta n_{k} + 	\delta n_{k} \mathcal{T}^{+}_{kp} \mathcal{T}^{-}_{pk} }{2} \right).
\end{align}
\noindent
Here $\mathcal{T}^{+}_{kp}\equiv\langle k | \mathcal{T}(\epsilon_k + i0^+)| p \rangle$ ($\mathcal{T}_{pk}^{-}$) is the retarded (advanced) on-shell $T$-matrix of a single impurity located at the origin. For graphene, $|k\rangle$ and $
|p\rangle$ are Bloch states of pristine graphene corresponding to the conduction band, for electron doping (valence band, for hole doping).

We stress that $\delta n_{k}$ and $\mathcal{T}_{kp}^{\pm}$ are matrices in spin space, so the order of factors in Eq.~\eqref{eq:coll-int} is important. The first term behaves as an effective (impurity-generated and momentum-dependent) magnetic field which respects time-reversal invariance. This ``magnetic field'' is precisely the hermitian part of the self-energy correction that can be derived in a diagrammatic calculation:
\begin{equation}\label{eq:self-energy}
\Sigma^R_{k}= \frac{ n_{\mathrm{imp}}}{2}(\mathcal{T}_{kk}^{+}+\mathcal{T}_{kk}^{-}).
\end{equation}
For impurity potentials that do not act upon the spin degree of freedom, this term has no dynamical consequences and can be absorbed by redefining the chemical potential. However, this is not the case for our problem because the $T$-matrix acts upon the spin degree of freedom. Note that a uniform 
Rashba coupling that  arises from encapsulation \cite{gumar2016spin} or an out-of-plane applied electric field~\cite{zibo2016numerical} will add an energy-independent potential to Eq.~\eqref{eq:self-energy}. Finally, the last two terms in Eq.~\eqref{eq:coll-int} are the quantum analogues of the ``scattered-in'' and ``scattered-out'' terms in the semiclassical Boltzmann equation.  In the first Born approximation, $\mathcal{T}^{\pm}_{kp} \to  V_{kp}$ (where $V$ is the single-impurity potential), the collision integral $\mathcal{I}[\delta n_k]$ would reduce to the more familiar form found in Refs.~\onlinecite{Glazov20102157,tarasenko2006scattering,pikus1991spin}. 

Next, note that the two $T$-matrices are related by hermitian conjugation:
\begin{equation}
\mathcal{T}_{pk}^{-} = \left[ \mathcal{T}^{+}_{kp}\right]^{\dag}.
\end{equation}  
Upon using the following parametrization: $\delta n_{k}=\rho_{k} \: \mathbb{1} + \boldsymbol{m}_{k}  \cdot \boldsymbol{\sigma}$, where $\rho_{k}$ and $\boldsymbol{m}_{k}$ represent the charge and spin distribution functions, respectively,  and $\mathcal{T}^{+}_{kp}=A_{kp} \: \mathbb{1} + \boldsymbol{B}_{k p} \cdot \boldsymbol{\sigma}$, we obtain:
\begin{align} 
\label{eq:QBE_charge}
&\partial_t \rho_k + e\boldsymbol{E} \cdot \frac{\boldsymbol{\nabla}_{k} n^{0}_{k}}{\hbar}  = \mathcal{I}_{1}[ \rho_{k}, \boldsymbol{m}_{k}], \\ 
 \label{eq:QBE_spin}
&\partial_t \boldsymbol{m}_k +  \left(\frac{\gamma}{\hbar} \boldsymbol{\mathcal{H}} - \frac{ n_{\mathrm{imp}}}{\hbar}\mbox{Re}\: \boldsymbol{B}_{kk}\right) \times \boldsymbol{m}_{k}
=\boldsymbol{\mathcal{I}}_{2}[\rho_{k},\boldsymbol{m}_{k}].
\end{align}
Note that the term involving $\boldsymbol{B}_{kk}$ in Eq.~\eqref{eq:QBE_spin} is related to the self-energy correction $\Sigma_{k}^{R}$;  This term is moved to the left-hand side of the equation to emphasis the resemblance between the impurity SOC-generated magnetic field $\boldsymbol{B}_{kk}$ and the real magnetic field $\boldsymbol{\mathcal{H}}$. The collision terms on the right of Eqs.~\eqref{eq:QBE_charge} and \eqref{eq:QBE_spin} describe how charge and spin are scattered by (time-reversal invariant) impurities; they are given by the following:
\begin{align}
\mathcal{I}_{1}[\rho_{k},\boldsymbol{m}_{k}] = \frac{n_{\mathrm{imp}}}{2\pi\hbar}\int\! & d^2 p \,
\bigg[ c_1 (\rho_{p}-\rho_{k}) + \boldsymbol{c}_2 \cdot(\boldsymbol{m}_{p}-\boldsymbol{m}_{k}) \nonumber \\
& - \boldsymbol{c}_3 \cdot(\boldsymbol{m}_{p}+\boldsymbol{m}_{k}) \,\bigg]\delta(\epsilon_p -\epsilon_k), \label{eq:I1} \\
\boldsymbol{\mathcal{I}}_{2}[\rho_{k},\boldsymbol{m}_{k}] = \frac{n_{\mathrm{imp}}}{2\pi\hbar} \int\! & d^2 p \,
\bigg[ c_1 (\boldsymbol{m}_{p}-\boldsymbol{m}_{k}) + \boldsymbol{c}_2 (\rho_{p}-\rho_{k}) \nonumber \\
&  +\boldsymbol{c}_3 \,(\rho_{p}-\rho_{k})
+ \boldsymbol{\mathcal{K}}\, \bigg]\delta(\epsilon_p -\epsilon_k), \label{eq:I2}
\end{align}
where the real-valued $c_1$, $\boldsymbol{c}_2$, $\boldsymbol{c}_3$ and $\boldsymbol{\mathcal{K}}$ are given by:
\begin{align}
  c_1 =& |A_{kp}|^{2}+|\mathbf{B}_{kp}|^{2} \; ; \; \boldsymbol{c}_2 = 2\text{Re}\: (A_{kp}\, \boldsymbol{B^{*}_{kp}}), \\
  \boldsymbol{c}_3 =& i\,\boldsymbol{B}_{kp}\times\boldsymbol{B}^{*}_{kp}, \\
    \boldsymbol{\mathcal{K}} =& 2  \mathrm{Im}(A_{kp}\boldsymbol{B}^{*}_{kp}) \times \boldsymbol{m}_{p} \nonumber
  +2\boldsymbol{B}^{*}_{kp}\times(\boldsymbol{B}_{kp}\times\boldsymbol{m}_{p}) \\ 
  &+ 2i\,\text{Im}\,[ (\boldsymbol{B}_{kp} \cdot\boldsymbol{m}_{p})\boldsymbol{B}^{*}_{kp} ]. \label{BigKappa}
\end{align}
The various terms in the collision integrals in Eqs.~\eqref{eq:I1} and \eqref{eq:I2} correspond to second-order scattering processes with specific physical interpretations [see Fig.~\ref{fig:triangle}(b)].  The $c_1$ terms describe conventional elastic scattering, and give rise to the Drude relaxation time.  The $\boldsymbol{c}_2$ terms give the skew scattering rate $\alpha_{\mathrm{sk}}$, which couples the charge current to the  spin current and is thus responsible for the extrinsic SHE.~\cite{Dyakonov_Perel_2}  The terms in $\boldsymbol{\mathcal{K}}$ contribute to the scattering rate induced by Rashba SOC $\alpha_{\mathrm{R}}$; the physical interpretation of these terms depends on the symmetry of the $T$-matrix and the dimensionality of the system. For a 3D electron gas with parity, rotational and time-reversal symmetry, the first (second) term in $\boldsymbol{\mathcal{K}} $ corresponds to the swapping spin current (EY spin relaxation) while the last term vanishes.~\cite{dyakonov_swapcurrent2009}  On the other hand, in a 2D non-relativistic electron gas with the same symmetry, $\boldsymbol{\mathcal{K}}$ gives rise to the extrinsic Edelstein effect, in addition to the swapping spin current and EY spin relaxation (see App.~\ref{app:sym}).

The $\boldsymbol{c}_3$ terms correspond to the ASP scattering mechanism [Fig.~\ref{fig:triangle}(c)]. For example, the terms involving $c_{3}^{y}$ contain the factors $(B_{kp})^x \sigma^x$  and 
$(B_{kp}^{*})^z \sigma^z$ which flips and skews the electron spin, respectively. Their product is proportional to $\sigma^y$ which  polarizes the electron spin in the $y$ direction. 
Note that the scattering process $\mathbf{c}_{2}$  couples to  $\boldsymbol{m}_{p}-\boldsymbol{m}_{k}$ in Eq.~\eqref{eq:I1} and cannot lead to a uniform magnetization. 

\section{Microscopic model and ansatz} \label{sec:model}

In order to solve the transport equations \eqref{eq:QBE_charge}--\eqref{BigKappa}, we require (i) a microscopic description \cite{hy2015extrinsic} of the scattering process that gives the single-impurity $T$-matrix $\mathcal{T}_{kp}^{\pm}$ , and (ii) an \emph{ansatz} for the distribution function \cite{chunli2016graphene}.

As mentioned above, the T-matrix is parameterized by $\mathcal{T}^{+}_{kp}=A_{kp} \: \mathbb{1} + \boldsymbol{B}_{k p} \cdot \boldsymbol{\sigma}$.  We can calculate $A_{kp}$ and $\boldsymbol{B}_{kp}$ for a microscopic model of 2D Dirac states scattering off an isolated rotationally and time-reversal symmetric impurity.  (Inter-valley scattering is neglected, because we are ultimately interested in impurities with characteristic size much larger than the inter-carbon distance in graphene.)  As shown in Appendix~\ref{app:micro_model}, the result is
\begin{align} 
\label{eq:t-matrix-A}
A_{kp}&= \gamma_{0}\cos\theta,\\
 \label{eq:t-matrix-B}
  \boldsymbol{B}_{kp}&= \Big[\gamma_{R} \sin\phi, \; -\gamma_{R} \cos\phi, \; i \gamma_{I} \sin\theta\Big],
\end{align}
\noindent
where $\theta \equiv (\theta_{k}-\theta_{p})/2$ and $\phi \equiv (\theta_{k}+\theta_{p})/2$, with $\theta_k \equiv \tan^{-1}(k_y/k_x)$ being the azimuthal angle for vector $k$. Thus, the results below will be expressed in terms of the (complex) renormalized potential strengths $\{\gamma_{0}, \gamma_I, \gamma_R\}$, which vary with the energy of the scattering electron (i.e.~chemical potential at zero temperature), and can exhibit resonances at certain energies. They are calculated from the bare impurity potentials which served as the inputs to the theory.  The bare impurity potentials can be fitted to \emph{ab initio} calculations and/or experiments (see Appendix~\ref{app:micro_model}). 

In order to obtain the steady state solution of the QBE, we introduce a drift velocity \emph{ansatz}  for the distribution function (which generalizes the one used in Ref.~\onlinecite{chunli2016graphene}):
\begin{multline} \label{eq:ansatz}
n^0_k  +\delta n_{k} \\= f_{\mathrm{FD}}\left[\epsilon_k - \hbar \boldsymbol{k} \cdot \boldsymbol{v}_{c} - ((\hbar\boldsymbol{k} \cdot \boldsymbol{v}_{s})\boldsymbol{\hat{n}}_{1} + h_0 \boldsymbol{\hat{n}}_{0})  \cdot \boldsymbol{\sigma}\right].
\end{multline}
Here, $n^0_k = f_{\mathrm{FD}}(\epsilon_k)$ is the equilibrium Fermi-Dirac distribution function, $\boldsymbol{v}_{c}$ ($\boldsymbol{v}_{s}$) is the drift velocity of the charge (spin) degrees of freedom, $h_0$ is proportional to the magnitude of the magnetization, and $\boldsymbol{\hat{n}_0}$ and  $\boldsymbol{\hat{n}_1}$ are the directions of magnetization and spin current polarization directly. The quantities of interest are the magnetization (i.e. non-equilibrium spin polarization), $\boldsymbol{M} = (M^x, M^y, M^z)$, the charge current density, $\boldsymbol{J} = (J_x, J_y)$, and  the spin current density $\boldsymbol{\mathcal{J}}^{a} = (\mathcal{J}_x^a, \mathcal{J}_y^a)$  (where $a=x,y,z$ is the spin orientation).
 At zero temperature, they are related with the \textit{ansatz} by 
\begin{align}
M^a =&\frac{\hbar g_{s}g_{v}}{\Omega}\sum_{k} (m_{k})^a  = \hbar g_{s}g_{\nu}N(\mu) h_{0} \: (\hat{n}_{0})^a,\\
J_i =& \frac{eg_{v} g_{s}}{\Omega}\sum_{k} \rho_{k} (v_{k})_{i} =eg_{s}g_{v}N(\mu)\epsilon_{F}\frac{(v_{c})_i}{2},\\
\mathcal{J}^{a}_i =&\frac{eg_{s}g_{v}}{\Omega}  \sum_{k}  (m_{k})^a (v_{k})_i =eg_{s}g_{v}N(\mu)\epsilon_{F}\frac{(v_{s})_{i} (\hat{n}_{1})^a }{2},
\end{align}
where $g_{s}=g_{v}=2$ are the spin and valley degeneracy, $\boldsymbol{v}_{k} = \hbar v_F (\mathbf{{k}}/k)$ is the group velocity, and $N(\mu)=\mu/(2\pi \hbar^2 v_{F}^2)$ is the density of states at Fermi energy. We follow the convention where spin current and charge current are measured in the same units.  For the sake of notational simplicity,  we have rescaled
\begin{equation}
  \boldsymbol{\mathcal{M}} = (ev_{F}/\hbar)\, \boldsymbol{M}.  
\end{equation}

Next, we substitute Eqs.~\eqref{eq:t-matrix-A}--\eqref{eq:ansatz} into Eqs.~\eqref{eq:QBE_charge}--\eqref{eq:QBE_spin} and set $\boldsymbol{E}=E_x$ and $\boldsymbol{\mathcal{H}}=0$.  This yields Eq.~\eqref{eq:transport}, the linear response relation.  The scattering rates that enter into this equation are given by:
\begin{align}
\alpha_{\mathrm{sk}}&= \frac{\pi n_{\mathrm{imp}} }{  \hbar}N(\mu)\: \mbox{Im}\left( \gamma_{I}\gamma_{0}^{*}\right),  \\
\alpha_{\mathrm{R}}&=  \frac{  n_{\mathrm{imp}} }{ \hbar } \left( \frac{1}{2}\mathrm{Re}\: \gamma_{R} + \pi N(\mu) \mathrm{Im}\:  (\gamma_{0}+\gamma_{I})\gamma_{R}^* \right), \label{eq:alpha_ext} \\
\alpha_{\mathrm{asp}}&= -\frac{2\pi n_{\text{imp}}}{	\hbar} N(\mu)\: \mbox{Re} \gamma_{I}\gamma_{R}^{*}, \label{eq:alpha_cf} \\
\frac{1}{\tau_{\mathrm{D}}}& =\frac{\pi n_{\mathrm{imp}}}{2\hbar}N(\mu)
\left( |\gamma_{0}|^2 + 3| \gamma_{I}|^2 + 4| \gamma_{R}|^2 \right), \\
\frac{1}{\tau_{\text{EY}} } &=  \frac{8}{\tau_{\mathrm{D}} }\left( \frac{|\gamma_{I}|^2+ |\gamma_{R}|^{2}}{  |\gamma_{0}|^2 + 3| \gamma_{I}|^2  + 4|\gamma_{R}|^2 } \right).
\end{align}
Note that $\alpha_{\mathrm{R}}$ contains a term linear in the SOC strength arising from forward scattering (i.e., the effective magnetic field induced by $\boldsymbol{B}_{kk}$). If we neglect Rashba SOC ($\alpha_{\mathrm{R}}=\alpha_{\mathrm{asp}}=0$) and set the magnetic field  $\boldsymbol{\mathcal{H}} = \boldsymbol{0}$, the spin precession vanishes and the QBE reduces to the semi-classical transport equation of Ref.~\onlinecite{ferreira2014extrinsic} (for the case of spin-conserving  SOC disorder), which captures the SHE but not CISP. Accounting for finite temperatures will complicate the expressions of the scattering rates without changing the results qualitatively.

Our theory neglects multiple impurity scattering and interband coherence effects. Therefore, it becomes less reliable when the Fermi energy $\mu$ becomes comparable to the temperature and, at zero temperature, when the Fermi wavelength is comparable to the average impurity distance. Using the impurity density $n_{\text{imp}}\sim 10^{10}\,\mathrm{cm}^{-2}$ reported in Ref.~\onlinecite{balakrishnan2014giant}, we estimate our theory breaks down for $\mu \approx 10$ meV.

\section{Summary and outlook} \label{sec:summary}

We have developed a quantum Boltzmann equation that exhibits three important technical advantages: (i) it accounts for the coherent spin dynamics of the electrons as they undergo scattering with a random ensemble of impurities that induced spin-orbit coupling by proximity; (ii) it goes beyond the standard Born approximation and treats impurity potential strength to all orders, thus capturing the important effect of scattering resonances; and (iii) it describes the experimentally-relevant dilute impurity regime (higher impurity densities can also be systematically accessed via Virial expansion if necessary; see  Appendix \ref{sec:app_QBE}).

Upon applying this theoretical framework to spin coherent transport in graphene with extrinsic SOC impurities, we found that besides the previously-known skew scattering and Rashba-induced spin flip scattering processes, there exists a distinct and novel scattering mechanism: anisotropic spin-precession (ASP) scattering. Conceptually, ASP scattering provides a ``missing link'' in the mechanisms linking spin polarization, charge current, and spin current, as shown diagrammatically in Fig.~\ref{fig:triangle}(a).  The most striking physical consequence is that ASP scattering provides a dominant contribution to current-induced spin polarization (CISP) in graphene.  This can be detected experimentally using either spatially resolved magneto-optical Kerr rotation~\cite{sih2005spatial} or suitable non-local transport measurements.  ASP scattering also gives a sizeable correction to the SHE, which could be verified by studying the spin current at low chemical potentials.  We have also calculated the spin relaxation time arising from SOC disorder, which includes both EY and DP relaxation.  The DP relaxation rate turns out to have a significant contribution from ASP scattering.  In the future,  it will be interesting to extend the transport equations to describe spin diffusion \cite{in-prep}, which may yield insights into ongoing experimental controversies over nonlocal resistance measurements in adatom-decorated graphene~\cite{neutral2015wang,kaverzin2015electron,balakrishnan2014giant}.

Apart from the graphene context, analogues of ASP scattering might also be present in other systems where 3D rotation symmetry is broken, such as interfaces between two different materials where roughness  \cite{sanchez2013spin} and impurities can generate SOC disorder.  The QBE formalism that we have developed can also be extended to study the anomalous Hall effect in ferromagnetic graphene, for which the anomalous Hall conductivity receives a large extrinsic contribution~\cite{wang2015proxmity}.

\textbf{Acknowledgements: } MAC's work is supported by the Ministry of Science and Technology (Taiwan) under contract number NSC 102-2112-M-007-024-MY5, and Taiwan's National Center of Theoretical Sciences (NCTS).  CH and CYD were supported by the Singapore National Research Foundation grant No.~NRFF2012-02, and by the Singapore MOE Academic Research Fund Tier 3 grant MOE2011-T3-1-005. We gratefully acknowledge useful discussions with  S.~Adam, E.~Farrell, R.~Raimondi, S.~Roche,  E. Sherman, J.~Sinova, S.~Valenzuela, and G.~Vignale.

\appendix

\section{Anisotropic spin precession scattering in two dimensional systems} \label{app:sym}

In this Appendix, we describe the generic form of the $T$-matrix of 2D electrons scattering off a spin-orbit coupling potential which exhibits rotational invariance (around the axis normal to the plane), time-reversal invariance and (in-plane) parity invariance. The goal is to show that the anisotropic spin precession (ASP) scattering term in the collision integral is in general non-vanishing for $T$-matrices with the above symmetries.

\subsection{Symmetry constraints on the $T$-matrix of two dimensional electron gas}

For simplicity, we begin our discussion with two dimensional electron gases (2DEG). The discussion below can be readily generalized to the case of graphene by accounting for Berry's phase. The $T$-matrix in the collision integral
of the Boltzmann transport equation is usually expressed in plane-wave
eigenstates $\langle \boldsymbol{r}| \boldsymbol{p}\rangle=\frac{1}{\sqrt{\Omega}}e^{i \boldsymbol{p} \cdot \boldsymbol{r}}$, where $\Omega$ is the 
area of the sample. In order to extract the angular dependence, it is more convenient to transform the $T$-matrix onto the (orbital) angular momentum basis
$ \langle r |E,m\rangle=2\pi J_{m}(kr)e^{im\theta}$:
\begin{align} \label{eq:T_temp}
\mathcal{T}_{pk}=& \,\langle \boldsymbol{p} |T(\boldsymbol{r})| \boldsymbol{k} \rangle \nonumber \\
=&\sum_{mm'}\langle m,E|T(r)|m',E\rangle\langle p |m,E\rangle\langle  k |m',E\rangle^{*}.
\end{align}
Note that the rotational invariance immediately implies that the impurity potential must be isotropic in 
space. The overlap between momentum basis and the (orbital) angular momentum basis is as follow:
\begin{eqnarray}
\langle\boldsymbol{p}|m,E\rangle & = & \int d^{2}r2\pi J_{m}(kr)e^{im\theta_{r}}e^{ipr\cos(\theta_{r}-\theta_{p})}\nonumber \\
  & = & c_m(E)e^{i m \theta_{p}},
\end{eqnarray}
where  
\begin{equation}
c_{m}(E)\propto\int rdr2\pi J_{m}(pr)2\pi i^{m}(pr)^{m}J_{m}(pr).
\end{equation}
Here $\theta_p = \arccos (\vec{\hat{p}}\cdot \vec{\hat{x}})$ is the angle subtended from the $x$-axis. With this, the $T$-matrix in the new basis is as follow:
\begin{align}
\mathcal{T}_{pk}=&\sum_{mm'}c_{m}(E)c_{m'}^{*}(E)\langle m,E|T(r)|m',E\rangle e^{i(m\theta_{p}-m'\theta_{k})}.
\end{align}
In what follows, it is convenient to project the $T$-matrix onto the unit and Pauli matrices:
\begin{equation}
T(r)=a(r) \mathbb{1} +b(r)s_{z}+\alpha(r)s^{+}+\beta(r)s^{-}.
\end{equation}
where $s^{\pm}=s_{x}\pm is_{y}$ and $a,b,\alpha,\beta$ are unknown functions of the position operator. For electron  states with spin aligned in the $z$ direction, the terms proportional to $a$ and $b$ are the spin conserving part of the $T$-matrix whereas terms proportional to $\alpha$ and $\beta$  are the spin-flip part of the $T$-matrix.

\subsubsection*{Rotational symmetry}
The $T$-matrix is invariant under rotation generated by the total angular
momentum $J_{z}=L_{z}+\frac{1}{2}s_{z}$, i.e.,
\begin{align}
& \,\langle m,E|T(r)|m',E\rangle  \nonumber \\
= & \langle m,E|e^{i(L_{z}+\frac{1}{2}s_{z})\chi}T(r)e^{-i(L_{z}+\frac{1}{2}s_{z})\chi}|m',E\rangle  \nonumber \\ 
= & e^{i(m'-m)\phi}\langle m,E|e^{i\frac{\chi}{2}s_{z}}T(r)e^{-i\frac{\chi}{2}s_{z}}|m',E\rangle
\end{align}
where $\chi$ is an arbitrary angle. The spin conserving part of the $T$-matrix commutes with the rotation
generated by $s_{z}$. For arbitrary angle $\chi$, the above condition
can only be fulfilled with $m=m'$. Hence,
\begin{align}
\langle p |a(r)| k \rangle &=\sum_{m}a_{m}(E)e^{im(\theta_{p}-\theta_{k})} \\
\langle p |b(r)s_{z}| k \rangle&=\sum_{m}b_{m}(E)e^{im(\theta_{p}-\theta_{k})}s_{z}
\end{align}
where $a_{m}(E)=|c_{m}(E)|^{2}\langle m,E|a(r)|m,E\rangle$ and $b_{m}(E)=|c_{m}(E)|^{2}\langle m,E|b(r)|m,E\rangle$. 
For the spin-flip part of the $T$-matrix, 
\begin{align}
& \, \langle m,E|\alpha(r)s^{+}|m',E\rangle   \nonumber \\
= & e^{i(m'-m)\chi}\langle m,E|\alpha(r)\left(e^{-\frac{i}{2}s_{z}\chi}s^{+}e^{\frac{i}{2}s_{z}\chi}\right)|m',E\rangle   \\ 
=  & e^{i(m'-m)\chi}\langle m,E|\alpha(r)\left(e^{-i\chi}s^{+}\right)|m',E\rangle\\
= & e^{i(m'-m-1)\chi}\langle m,E|\alpha(r)|m',E\rangle s^{+}
\end{align}
For arbitrary angle $\chi$, the equality above can only be satisfied
by $m'=m+1$. Similar arguments applied to $\beta(r)s^{-}$, therefore
\begin{align}
\langle p|\alpha(r)s^{+}  k \rangle &=\sum_{m}\alpha_{m}(E)s^{+}e^{im\theta_{p}}e^{-i(m+1)\theta_{k}}  \\
\langle p|\beta(r)s^{-}| k \rangle &=\sum_{m}\beta_{m}(E)s^{-}e^{i(m+1)\theta_{p}}e^{-im\theta_{k}}
\end{align}
with $\alpha_{m}(E)=\langle m,E|\alpha(r)|m+1,E\rangle c_{m}(E)c_{m+1}^{*}(E)$
and $\beta_{m}(E)=\langle m,E|\beta(r)|m+1,E\rangle c_{m+1}(E)c_{m}^{*}(E)$.
The physics of these equations is clear: Since the total angular momentum
is conserved, an increment in the spin angular momentum is compensated
by a decrement in the orbital angular momentum. In summary, using
rotational invariance, the $T$-matrix is constrained to the following form:
\begin{align}
\mathcal{T}_{kp}=
\sum_{m}\big( & a_{m}(E)+b_{m}(E)s_{z} +\alpha_{m}(E)e^{-i\theta_{k}}s^{+} \nonumber \\
&  +\beta_{m}(E)e^{i\theta_{p}}s^{-}\big)e^{im(\theta_{p}-\theta_{k})}.
\end{align}
\subsubsection*{Time-reversal symmetry}
To further constrain the $T$-matrix, we invoke time-reversal symmetry. Under time reversal, the matrix elements of the $T$-matrix undergoes the following change:
\begin{equation}
\langle\phi|T|\psi\rangle\rightarrow\langle\Theta_{T}\psi|T|\Theta_{T}\phi\rangle
\end{equation}
where $\Theta_{T}$ is the time-reversal operator. 
Effectively, this would map $\theta_{p}\rightarrow\theta_{k}+\pi$, $\theta_{k}\rightarrow\theta_{p}+\pi$, $s^{\pm}\rightarrow-s^{\pm}$ and $s_{z}\rightarrow-s_{z}$.

Under time reversal invariance, the scalar part of the $T$-matrix must
be an even function of the scattering angle $\theta \equiv \theta_{p}-\theta_{k}$;
the $s_{z}$ part of the $T$-matrix must be an odd function of the scattering
angle  $\theta \equiv \theta_{p}-\theta_{k}$. Hence, the spin-conserving part
of the $T$-matrix must be of the form

\begin{equation}
\sum_{m}a_{m}(E)\cos m(\theta_{p}-\theta_{k})+b_{m}(E)s_{z}\sin m(\theta_{p}-\theta_{k}).
\end{equation}

The spin flip part of the $T$-matrix can be written as the following,
\begin{align}
\sum_{m} \big( &\alpha_{m}(E)e^{-i\left(\frac{\theta_{k}+\theta_{p}}{2}\right)}e^{i(m+\frac{1}{2})(\theta_{p}-\theta_{k})}s^{+} \nonumber \\
&+ \beta_{m}(E)e^{i\left(\frac{\theta_{k}+\theta_{p}}{2}\right)}e^{i(m+\frac{1}{2})(\theta_{p}-\theta_{k})}s^{-}\big)
\end{align}
Under time-reversal invariance the sum of the angle $\phi\equiv\theta_{k}+\theta_{p}\rightarrow\phi+2\pi$
. Hence, the spin flip part of the $T$-matrix must be an even function
of the  scattering angle:
\begin{align}
\mathcal{T}_{kp}=&\sum_{m}a_{m}(E)\cos (m\theta) + b_{m}(E)\sin (m\theta) \, s_{z} \nonumber \\
&+ \left(s^{+}\alpha_{m}(E)e^{-i\frac{\phi}{2}}+s^{-}\beta_{m}(E)e^{i\frac{\phi}{2}}\right)\cos\left( \frac{(2m+1)\theta)}{2}\right)\label{eq:t-matrix}
\end{align}
where,
\begin{equation} 
\theta=\theta_{p}-\theta_{k}\;;\;\phi=\theta_{p}+\theta_{k} \label{eq:sum_diff}
\end{equation}

\subsubsection*{Parity  symmetry}

Parity symmetry in two dimension system acts like an mirror symmetry. Under mirror symmetry about the $yz$ plane, $\{x,y,z\}\rightarrow\{-x,y,z\}$
and $\{s_{x},s_{y},s_{z}\}\rightarrow\{s_{x},-s_{y},-s_{z}\}$. Effectively, this maps $s^{\pm}\rightarrow s^{\mp}\;;s_{z}\rightarrow-s_{z}$ , $\theta_{k}\rightarrow\pi-\theta_{k}\;;\;\theta_{p}\rightarrow\pi-\theta_{p}$ and $\theta\rightarrow-\theta\;;\;\phi\rightarrow2\pi-\phi$.
After rotational and time-reversal symmetry, the parity symmetry does
not put any further constraint on the spin-conserving part of the $T$-matrix. It
does, however, constraint the spin flip part: $\alpha_{m}(E)=-\beta_{m}(E)$. This leads us to the form of Rashba spin orbit coupling:
\begin{align}
\mathcal{T}_{kp} =& \sum_{m}a_{m}(E)\cos m\theta+s_{z}b_{m}(E)\sin m\theta \nonumber \\
& \quad +\alpha_{m}(E)\left(s^{+}e^{-i\frac{\phi}{2}}-s^{-}e^{i\frac{\phi}{2}}\right)
\cos\left((m+\frac{1}{2})\theta\right)\nonumber \\
= & \sum_{m}a_{m}(E)\cos m\theta + s_{z}b_{m}(E)\sin m\theta \nonumber \\
 &\quad +\alpha_{m}(E)\left(s_{x}\sin\frac{\phi}{2}-s_{y}\cos\frac{\phi}{2}\right)\cos\left((m+\frac{1}{2})\theta\right)  , 
\end{align}
where in the last line we have further defined $\alpha_{m}(E)\rightarrow-2i\alpha_{m}(E)$.
The peculiar half angle dependence of the spin flip part of the $T$-matrix
arises from the rotational invariance in two dimensions. 
It is more convenient to restore the angular dependence and write the $T$-matrix in the following form:
\begin{equation} 
\mathcal{T}_{kp} \equiv a(E,\theta) + s_{z} b(E,\theta) +\alpha(E,\theta) \left(s_{x}\sin\frac{\phi}{2}-s_{y}\cos\frac{\phi}{2}\right).  \label{eq:t-matrix_full}
\end{equation}
Keeping in mind that $ a(E,\theta)$ and $\alpha(E,\theta)$ is an even function of $\theta$; $ b(E,\theta)$ is an odd function of $\theta$.

\subsection{Symmetry constraints on the $T$-matrix of graphene }

For doped graphene, instead of projecting the $T$-matrix onto the plane wave eigenstates, we project it onto the conduction (valence) band eigenstates $\langle  \mathbf{p} \pm|T|   \mathbf{k} \pm\rangle$ where ($+$ applies to electron doping and $-$ to hole doping):
\begin{equation}
\langle \boldsymbol{r} | \boldsymbol{p} \pm\rangle = 
\frac{1}{\sqrt{2}}\left(\begin{array}{c}
e^{-i \theta_p /2 }  \\
\pm e^{i \theta_p /2 } \\
 \end{array}\right)e^{i  p \cdot  r}.
\end{equation}
We recall inter-band transitions (i.e. $\langle  \boldsymbol{p} +|T|  \boldsymbol{k} -\rangle$  and $\langle  \boldsymbol{p} -|T|  \boldsymbol{k} +\rangle$ ) are forbidden for elastic scattering.  For impurities with characteristic size $R \gg a$, where $a=0.25 $nm is the distance between carbon atoms in graphene, inter-valley scattering in graphene can be neglected. Hence, following the same derivation we did for the 2DEG,
we can arrive at the same equation,
\begin{equation}
\mathcal{T}_{kp} = a(E,\theta) + s_{z} b(E,\theta) +\alpha(E,\theta) \left(s_{x}\sin\frac{\phi}{2}-s_{y}\cos\frac{\phi}{2}\right)  
\end{equation}
where $ a(E,\theta)$ and $\alpha(E,\theta)$ is an even function of $\theta$; $ b(E,\theta)$ is an odd function of $\theta$ (recall that $\theta\equiv \theta_k -\theta_p$ is the scattering angle). However, note that time-reversal symmetry in graphene affects the valley degree of freedom, $K \rightarrow K'$ and $K' \rightarrow K$. Since inter-valley scattering is neglected and we are only interested in the $T$-matrix for single valley, the ``valley time-reversal symmetry'' implemented by $\Theta_{s}= (-i s_y \mathcal{K})\sigma_x \tau_x $ is used to constraint the $T$-matrix instead of the (total) time-reversal symmetry. Here $\mathcal{K}$ is the anti-unitary operator; $\sigma$, $\tau$, $s$ and represent the sublattice pseudospin, valley pesudospin and real electron spin respectively. The valley time reversal symmetry consists of a time-reversal symmetry, followed by inversion and a $\pi$ rotation about the axis perpendicular to graphene plane.

In spite of the similarities between the $T$-matrix for graphene and 2DEG, there is one crucial difference arising from the Berry phase; when $\theta_p \to \theta_p + 2\pi$ and $\theta_k \to \theta_k$ (hence
$\theta \to \theta + 2\pi$, and $\phi\to \phi+ 2\pi$), the $T$-matrix of graphene changes sign. Therefore,
\begin{align}
a(E,\theta + 2\pi) = -a(E,\theta),\\
b(E,\theta+2\pi) = -b(E,\theta),\\
\alpha(E,\theta + 2\pi) = +\alpha(E,\theta).
\end{align}
Note that in a 2DEG, the $T$-matrix has the same structure
but there is no (sublattice) pseudo-spin and therefore there is no change of sign  associated with the Berry phase:
\begin{align}
a(E,\theta + 2\pi) = a(E,\theta),\\
b(E,\theta+2\pi) = b(E,\theta),\\
\alpha(E,\theta + 2\pi) = - \alpha(E,\theta).
\end{align}
In summary, although the $T$-matrix of \emph{doped} graphene and 2DEG are formally identical, the periodicity of the $T$-matrix parameters are different, due to the existence of the Berry phase. For graphene, $a(\theta)$  and $b(\theta)$ are $4\pi$ periodic, i.e. functions of $\theta/2$, whereas $\alpha(\theta)$ is $2\pi$ periodic, i.e. a function of $\theta$. On the other hand, for 2DEG, $a(\theta)$ and $b(\theta)$ are functions of $\theta$ with periodicity $2\pi$, and $\alpha(\theta)$ is a function of $\theta/2$ with  periodicity $4\pi$.

\subsection{Aniostropic spin precession scattering}

In the main text, we argued that the anisotropic spin precession (ASP) scattering can occur in general two-dimensional metallic systems with time-reversal symmetry, rotational symmetry and parity symmetry. We will verify the claim in this section with the $T$-matrices derived in the previous section. First, recall the ASP scattering term is described by the collision integral:
\begin{equation} \label{eq:app_asp}
I_1 =\frac{2\pi n_{\mathrm{imp}}}{\hbar}\int d^{2}p\; \boldsymbol{c}_{3}.(\boldsymbol{m}_{k}+\boldsymbol{m}_{p}) \delta(\epsilon_k -\epsilon_p)
\end{equation}
where
\begin{align}
 \boldsymbol{m}_{k}=&\frac{1}{2} \text{Tr} \:\left(\delta n_k\boldsymbol{\sigma} \right), \\
 \boldsymbol{c}_{3}=& i \,\boldsymbol{B}_{kp}\times\boldsymbol{B}^{*}_{kp}
\end{align}
From the previous discussion, it is clear that 
\begin{align}
A_{kp} &= a(\theta),\\
\boldsymbol{B}_{kp} &= \left[ \alpha(\theta) \sin \left( \frac{\phi}{2}\right)\,,\, -\alpha(\theta) \cos \left( \frac{\phi}{2} \right) \, , \, b(\theta)\right] \nonumber \\
&= \boldsymbol{B}^{\perp}_{kp} + B^{\parallel}_{kp} \hat{\boldsymbol{z}},
\end{align}
where,
\begin{align}
 \boldsymbol{B}^{\perp}_{kp} &= \alpha(\theta) \left[ \sin \left( \frac{\phi}{2} \right)\,,\,
-\cos \left( \frac{\phi}{2} \right) \,,\,0 \right],\\
 B^{\parallel}_{kp}  &=  b(\theta).  
\end{align}
The ASP scattering cross-section is then given by
\begin{align}
\boldsymbol{c}_{3} &= i \left( \boldsymbol{B}^{\perp}_{kp} + B^{\parallel}_{kp} \boldsymbol{\hat{z}} \right) \times \left( (\boldsymbol{B}^{\perp}_{kp})^* + (B^{\parallel}_{kp})^* \boldsymbol{\hat{z}} \right)  \nonumber \\
&=  i \boldsymbol{B}^{\perp}_{kp} \times  
 (\boldsymbol{B}^{\perp}_{kp})^* - 2  \mathrm{Im}\left[ \boldsymbol{B}^{\perp}_{kp} (B^{\parallel}_{kp})^* \right]\times \boldsymbol{\hat{z}}. \label{eq:app_c3}
\end{align}
However, we note that $\boldsymbol{B}^{\perp}_{kp} \propto \left(\boldsymbol{B}^{\perp}_{kp} \right)^*$ and therefore the first term vanishes. To make further progress, we substitute the generalized drift-velocity \textit{ansatz} \cite{chunli2016graphene} of the distribution matrix (Eq.~\ref{eq:ansatz}) into Eq.~\ref{eq:app_c3}. Then,
\begin{align}
\boldsymbol{c}_3 
= 2  \mathrm{Re}\left[ \alpha(\theta) b^*(\theta) \right] \boldsymbol{\hat{n}}(\phi).
\end{align}
where
$\boldsymbol{\hat{n}}(\phi)= \left( \cos \frac{\phi}{2}\,,\,
\sin \frac{\phi}{2}\,,\,0 \right)$.



In order to couple magnetization $\boldsymbol{\mathcal{M}}$ to charge current  $\boldsymbol{J}$, the following integral must not vanish:
\begin{equation}
\boldsymbol{J} \propto  \int d^2 k  \int d^2 p \, \boldsymbol{v}_k   \,
\delta(\epsilon_k -\epsilon_p) \mathbf{c}_3 
\cdot (\boldsymbol{m}_{k}+\boldsymbol{m}_{p})  .
\end{equation}
Since magnetization corresponds to the zeroth harmonic mode oscillation of the Fermi surface, we further expand $\vec{m}_k = \delta(\epsilon_k -\mu) \vec{h}_0 + ...$. The zeroth harmonic mode $\vec{h}_0=h_0 \vec{n}_0$ is the magnetization and higher harmonic modes are not important in the discussion of the ASP effect (see Eq.~\eqref{eq:ansatz}). The integration can be further simplified into the following:
\begin{equation}
\boldsymbol{J} \propto  \int d\theta_k  \int d\theta_p \, \hat{\vec{k}} \,  \mathrm{Re}\left[ \alpha(\theta) b^*(\theta) \right] (\boldsymbol{\hat{n}}(\phi)
\cdot 2\vec{h}_0 )  .
\end{equation}
Here $\hat{\vec{k}}=(\cos \theta_k , \sin \theta_k )$ determines the direction of the current. The integral can be simplified by expressing $\hat{\vec{k}}$ as follow:
\begin{align}
\hat{\vec{k}} = \cos\left( \frac{\theta}{2} \right) \vec{\hat{n}}(\phi) + \: 
\sin \left( \frac{\theta}{2}\right) \left(\vec{\hat{n}}(\phi) \times \vec{\hat{z}} \right),
\end{align}
where $\theta$ and $\phi$ are defined in Eq.~\ref{eq:sum_diff}.
Next, using the symmetries of the $T$-matrix discussed earlier (i.e.~ $\alpha(-\theta) = \alpha(\theta)$ and $b(-\theta) = -b(\theta)$), the term
proportional to $\cos\theta/2$ in the integrand vanish upon integration. Hence,
\begin{align}
\boldsymbol{J} \propto &
\int^{2\pi}_0 d\theta \: \mathrm{Re}\left[\alpha^*(\theta) b(\theta) \right]  \sin \left( \frac{\theta}{2}\right)\times \nonumber \\  
\,\quad & \int^{2\pi}_0 d\phi \, \left(\vec{\hat{n}}(\phi)\cdot \vec{h}_0 \right)  
\left(\vec{\hat{n}}(\phi) \times \vec{\hat{z}} \right).
\end{align}
Finally, we note that 
\begin{align}
 \int^{2\pi}_0 d\phi \, \left(\vec{\hat{n}}(\phi)\cdot \vec{h}_0 \right)   
\left(\vec{\hat{n}}(\phi) \times \vec{\hat{z}} \right)  
=  \vec{\hat{z}}\times \vec{h}_0.
\end{align}
Form which we arrive at the result:
\begin{equation}
\boldsymbol{J}
\propto \left(\vec{\hat{z}}\times \vec{h}_0 \right)  \int^{2\pi}_0 d\theta\: \mathrm{Re}\left[\alpha^*(\theta) b(\theta) \right] \: \sin \left( \frac{\theta}{2} \right) \: 
\end{equation}
The $\theta$ integral above is non-vanishing for general $\alpha(\theta)$ and $b(\theta)$. Note that the direction of the induced current is perpendicular to the magnetization. Hence, in the presence of an electric field in the $x$-direction, the current induced spin polarization (CISP) is non-vanishing the $y$-direction but CISP in the $x$-direction is not excited. 

In summary, using the parameters of a single impurity $T$-matrix with parity, time-reversal and rotational symmetry, we show that ASP scattering can indeed couple magnetization with charge current directly for 2DEG and graphene.

\section{Microscopic model of $T$-matrix in graphene} \label{app:micro_model}
In this section, we present a microscopic model of the $T$-matrix. The underlying hamiltonian is

\begin{equation} \label{eq:H+V}
\mathcal{H}=\hbar v_F\left(\pm\sigma_x p_x+\sigma_y p_y\right)+\sum_{\alpha=0,I,R} \lambda_{\alpha} R^2 \Lambda_{\alpha}\delta^{(2)}(\mathbf{r})
\end{equation} 
where $\lambda_{\alpha}$ is the strength of the potential, $R$ the radius of the scatterer ($\sim$ range of the potential) , and $\Lambda_{\alpha}$ are $4\times 4$ matrices acting upon the sublattice-spin degrees of freedom. Explicitly, the $\Lambda$ matrices are,
\begin{align}
\Lambda_{0}=\mathcal{I}_{4\times 4}, \; ;  \; \Lambda_{I}=\sigma_{z}s_{z},   \; ; \;
\Lambda_{R}= \pm \sigma_{x} s_{y}-\sigma_{y}s_{x} .
\end{align}
Here $\Lambda_0$ is the spin-independent scalar potential,  $\Lambda_I$ is the Kane-Mele type spin-conserving spin-orbit coupling and $\Lambda_R$ is the Rashba type spin-flip spin-orbit coupling. The bare potential strengths serve as the inputs into our theory and they can be extracted from numerical and experimental studies. The $\pm$ stands for the two valleys in graphene. In this model, we have assumed that the disorder potential varies very slowly on the lattice scale of graphene and cannot induce inter-valley scattering. This model can be solved exactly \cite{hy2015extrinsic} and gives the following $T$-matrix:
\begin{align}
\mathcal{T}_{kp} =& \gamma_{0}(k) \cos \left(\frac{\theta_k -\theta_p}{2}\right) + i \gamma_{I}(k)\sin \left(\frac{\theta_k -\theta_p}{2} \right) s_{z}  \nonumber \\
 &+\gamma_{R}(k) \left( \sin \left(\frac{\theta_k +\theta_p}{2}\right) s_{x}    -  \cos \left( \frac{\theta_k +\theta_p}{2} \right) s_{y}   \right). 
\end{align}
Here $(\theta_{k}-\theta_{p})/2$ and $(\theta_{k}+\theta_{p})/2$ reflect the Berry phase of the $T$-matrix. That is, rotating the scattered state by $2\pi$, (i.e.  $\theta_p\rightarrow \theta_p + 2\pi$) while keeping the incident state $\theta_k$ unchanged, the $T$-matrix gets back to itself with an extra minus sign. The $\gamma(k)$'s are complex numbers and they represent the renormalized potential strength as a function of incident energy. They are related with the bare impurity potential by the following equations:

\begin{widetext}
\begin{align}
\gamma_0 (k)&= \frac{1}{4G_0(k)} \left( \frac{1}{ G_{0}(k) (-\lambda_0+ \lambda_I -2  \lambda_R)+1} +
\frac{1}{G_{0}(k)  (-\lambda_0 + \lambda_I +2\lambda_R)+1}-\frac{2}{G_{0}(k)
   (\lambda_0 +\lambda_I)-1}-4 \right), \\
\gamma_I (k) &= \frac{\lambda_{I}+\lambda_{I}G_{0}(k)(\lambda_{I}-\lambda_{0})-2G_{0}(k)\lambda_{R}^{2}}{\left(1-G_{0}(k)(\lambda_{I}+\lambda_{0})\right)\left(1-G_{0}(k)(\lambda_{0}-\lambda_{I}-2\lambda_{R})\right)\left(1-G_{0}(k)(\lambda_{0}-\lambda_{I}+2\lambda_{R}))\right)} ,\\
\gamma_R (k) &= \frac{\lambda_{R}}{\left(1+G_{0}(k)(\lambda_{I}-\lambda_{0})\right)^{2}-4G_{0}^{2}(k) \lambda_{R}^{2}},
\end{align}
where
\begin{align}
G_0(k)=\text{sign}(E)\frac{k}{2\pi\hbar v_F}\log |kR|-\frac{ik}{4\hbar v_F},
\end{align}
\end{widetext}
is the Green function at the origin. It is  obtained by imposing a cut-off at  momentum $k^{\prime}
\sim R^{-1}$. In the main text, results are plotted for the strong Rashba and weak Rashba cases. Strong Rashba SOC corresponds to $\lambda_R  =30$~meV whereas weak Rashba SOC corresponds to  $\lambda_R =10$~meV . In both situation, $\lambda_0 =80$~meV and  $\lambda_I =10$~meV. These  parameters are consistent with the range of parameters provided by \emph{ab initio} \cite{netoguinea09,weeks2011engineering,stabilizing2012hua,fabian2013spin,fabian2015spin} or experimental  \cite{marchenko2012giant,calleja2015natphy,balakrishnan_colossal,balakrishnan2014giant}  studies. However, note that the exact values of the bare potential strengths depend on the microscopic details of a particular graphene sample.

\begin{widetext}
\section{Derivation of the Quantum Boltzmann Equation}
\label{sec:app_QBE}
In this appendix, we discuss in detail the derivation of the Quantum Boltzmann Equation (QBE) starting from the quantum Liouville equation, following the methods pioneered by Kohn and Luttinger \cite{KohnLuttingerBTE}. The derivation of the QBE consists of three steps. First, the linearized quantum Liouville equation (or linearized QBE) is transformed into two operator equations by introducing the collision integral $\Gamma_{k}$. 
The two operator equations correspond to the diagonal (in momentum) part of the QBE Eq.~\eqref{eq:nk}, and off-diagonal  (in momentum) part of the QBE Eq.~\eqref{eq:geq}. Then, we eliminate (i.e. integrate out) the off-diagonal part of the QBE in favor of the diagonal QBE to arrive at a self-consistent equation for $\Gamma_{k}$, Eq. \eqref{eq:gamma1}. Lastly, a Virial (power) expansion on the impurity density is used to solve the self-consistent equation for  $\Gamma_{k}$ and we arrive at the QBE.
\subsection{Definition of single particle density operator}
Before discussing the details of the QBE, we made a small digression to discuss the one-particle density matrix operator. We define the one particle density operator as follows
\begin{equation}
\rho(t)  = \sum_{kp \alpha  \beta} |k \alpha \rangle \: \langle c^{\dag}_{p \beta}(t) c_{k \alpha}(t) \rangle \: \langle p \beta  |.\label{eq:densop1}
\end{equation}
Here $c_{p\alpha}$ ($c_{p\alpha}^{\dagger}$) is the annihilation (creation) operator of momentum $p$ and quantum number $\alpha$. In this notes, the vector notation for momentum $\vec{p}$ and coordinate $\vec{r}$ are denoted by $p$ and $r$ respectively. In the context of the Boltzmann transport equation for electrons, the annihilation and creation operators obey the fermionic anticommutation relationship $\{c_{p\alpha},c_{k\beta}^{\dagger} \}=\delta_{kp}\delta_{\alpha\beta}$. The thermodynamic average is given by the following equation,
\begin{align}
\langle c^{\dag}_{p \beta} (t) c_{k \alpha}(t) \rangle &= \frac{1}{Z}\: 
\mathrm{Tr} \left[e^{-(H-\mu N)/k_B T}U^{\dag}(t) c^{\dag}_{p \beta} c_{k \alpha} U(t) \right] \nonumber \\ 
&=  \mathrm{Tr} \left[ c_{k\alpha}  U(t) \frac{e^{-(H-\mu N)/k_B T}}{Z} U^{\dag}(t) c^{\dag}_{p \beta} \right] \nonumber\\
&= \mathrm{Tr}\left[c_{k\alpha} \Lambda(t) c^{\dag}_{p\beta}\right],\label{eq:dens2}
\end{align}
where $Z = \mathrm{Tr} \: e^{-(H-\mu N)/k_B T}$
is the partition function in the grand canonical ensemble (at absolute temperature $T$ and chemical potential $\mu$) and $U(t)$ is the evolution operator which obeys the equation:
\begin{equation}
i\hbar \partial_{t} U(t) = H(t) U(t)
\end{equation}
with $H(t)$ being the Hamiltonian of the system at time $t$. 
The operator
\begin{equation}
\Lambda(t) = U(t) \frac{e^{-(H-\mu N)/k_B T}}{Z} U^{\dag}(t)
\end{equation}
is the many-particle density matrix at time $t$. 
Note that this operator obeys the equation:
\begin{equation}
i\hbar \partial_t \Lambda(t) = \left[ H(t), \Lambda(t) \right] .
\end{equation}
The strange index assignment of the single-particle density matrix becomes more clear  if we use Eq.~\eqref{eq:dens2} as well as the cyclic property of the trace and rewrite it as:
\begin{equation}
\rho^{\alpha\beta}_{k p}(t) = \langle k \alpha | \rho(t) | p \beta \rangle = \langle c^{\dag}_{p\beta}(t) c_{k\alpha}(t) \rangle =   \mathrm{Tr}\left[c_{k\alpha} \Lambda(t) c^{\dag}_{p\beta}\right]
\end{equation}

For a system of non-interacting particles, the Hamiltonian can be written as follows:
\begin{equation}
H(t) = \sum_{kp\alpha\beta} \left[\epsilon_{k}^{\alpha\beta}   \delta_{kp} + u^{\alpha\beta}_{kp}(t) \right]  c^{\dag}_{k\alpha} c_{p\beta} =
\sum_{kp\alpha\beta} h^{\alpha\beta}_{kp}(t)   c^{\dag}_{k\alpha} c_{p\beta},
\end{equation}
where $\epsilon_k$ are the energies of the Bloch orbitals  and $u^{\alpha\beta}_{kp}$ describes a time and spin-dependent potential. In its first quantized form the single-particle Hamiltonian can be written as
\begin{equation}
h(t) = h_0 + u(t),
\end{equation}
where,
\begin{align}
h_0 &= \sum_{k\alpha\beta} \epsilon^{\alpha\beta}_k |k \alpha \rangle \langle k \beta |, \\
u(t) &= \sum_{k p\alpha\beta} u^{\alpha\beta}_{kp} (t)
|k \alpha\rangle \langle p \beta|, \\
h(t) &= \sum_{kp\alpha\beta} h^{\alpha\beta}_{kp}(t)\: 
|k \alpha\rangle \langle p \beta| .
\end{align}
In what follows, we shall derive the equation of motion for the single-particle density matrix operator $\rho(t)$ introduced in Eq.~\eqref{eq:densop1}. To this end, we need to use the equations of motion for the operator $c^{\dag}_{p\beta}(t) c_{k\alpha}(t)$:
\begin{align}
&i \hbar \partial_t c^{\dag}_{p\beta}(t)c_{k\alpha}(t) \nonumber \\ &= \left[ c^{\dag}_{p\beta}(t) c_{k\alpha}(t), H(t) \right] \nonumber \\
 &= -\sum_{l\gamma}  h^{\gamma \beta}_{lp}(t) c^{\dag}_{l\gamma}(t) c_{k\alpha}(t) + \sum_{q\delta} h^{\alpha\delta}_{kq}(t) c^{\dag}_{p\beta}(t)c_{q\delta}(t)  \nonumber \\
 &=  \sum_{l\gamma} \left[   h^{\alpha\gamma}_{kl}(t) c^{\dag}_{p\beta}(t)c_{l\gamma}(t) - h^{\gamma \beta}_{lp}(t) c^{\dag}_{l\gamma}(t) c_{k\alpha}(t) \right],
\end{align}
where we have used
\begin{align}
\left[ c^{\dag}_{l\gamma} c_{q\delta}, c^{\dag}_{p\beta} c_{k\alpha} \right] &= c^{\dag}_{l\gamma} \left[  c_{q\delta}, c^{\dag}_{p\beta} c_{k\alpha}\right] + \left[ c^{\dag}_{l\gamma},
 c^{\dag}_{p\beta} c_{k\alpha} \right]c_{q\delta} \nonumber \\
&=  \delta_{qp}\delta_{\delta\beta} c^{\dag}_{l\gamma} c_{k\alpha} - \delta_{lk} \delta_{\gamma\alpha} c^{\dag}_{p\beta}c_{q\delta}.
\end{align}
Hence, by taking the expectation value of this equation and using   $\rho^{\alpha\beta}_{kp}(t) = \langle c^{\dag}_{p\beta}(t) c_{k \alpha}(t) \rangle$, we arrive at:
\begin{equation}
i \hbar \partial_t \rho^{\alpha\beta}_{kp}(t) 
= \sum_{l\gamma} \left[ h^{\alpha \gamma}_{kl}(t) 
\rho_{lp}^{\gamma\beta}(t)  - \rho^{\alpha\gamma}_{kl}(t) h^{\gamma\beta}_{lp}(t)  \right], 
\end{equation}
which in operator form reduces to:
\begin{equation}
i\hbar \partial_t \rho(t)  = \left[ h(t), \rho(t) \right] = \left[ h_0 + u(t), \rho(t) \right]. 
\end{equation}
In summary, we have shown that the one-particle density operator defined in Eq.~\eqref{eq:densop1} obeys the quantum Lioville equation.
\subsection{Extension of the Kohn-Luttinger formalism} 
In this section, we shall develop an extension to the Kohn and Luttinger (henceforth referred to as KL) derivation of the Boltzmann transport equation to leading order in the density of impurities $n_{\mathrm{imp}}$~\cite{KohnLuttingerBTE}. This kinetic 
equation can be used to study spin and/or (pesudo-spin) and charge
transport in systems contaminated by a \emph{dilute},
random ensemble of  impurities for which spin coherence are important. The derivation presented below extends KL's work in two directions: 
First, the quantum coherence of the spin dynamics is taken into account. Second, we account for the possibility of time dependent electric and magnetic fields, whose characteristic frequency is much smaller than the Fermi energy, $\mu$.

We consider a non-interacting system where the single-particle Hamiltonian is of the form:
\begin{equation}
h_E(t) = h - e  \vec{E}(t)\cdot \vec{r} + \gamma  \boldsymbol{\mathcal{H}}(t)\cdot \boldsymbol{s},\qquad h = h_0 + u, 
\end{equation}
where $\boldsymbol{\mathcal{H}}(t) \sim  \vec{\mathcal{H}}(\omega) e^{-i \omega t}$ is a time-dependent external magnetic field and $\gamma$ is the gyromagnetic factor. $\vec{E}(t) \sim \vec{E}(\omega^{+}) e^{-i \omega^{+} t}$ is an AC electric field that drives a current through the system ($\omega^{+} = \omega + 2i \eta$, $\eta\to 0^+$,  which contains a small imaginary part, to account for the adiabatic switching  of the external perturbations). $u = \sum_{i=1}^{N_{\mathrm{imp}}} u_i$ is the sum of impurity potentials located at $\vec{r}_i$ and $h_0$ is  the spin and (crystal) momentum diagonal  part of the Hamiltonian,
\begin{equation}
h_0 = \sum_{k\alpha} \epsilon_{k} \: |k\alpha\rangle \langle k \alpha | 
\end{equation}
Following KL, for small $|\vec{E}(\omega)|$, we split the one-particle density matrix into two parts:
\begin{equation}
\rho_E(t)  = \rho + \delta \rho(t),
\end{equation}
where
\begin{equation}
\rho = \frac{1}{e^{(h-\mu)/k_B T} + 1}
\end{equation}
and  $\delta \rho(t) = \delta \rho(\omega) e^{-i\omega^{+} t}$,  $\delta \rho(\omega) \propto   |\vec{E}(\omega)|$. Thus, to leading order in 
$|\vec{E}(t)|$ the linearized quantum Liouville equation (LQLE) reads:
\begin{equation}
\left[h, \delta \rho(t) \right]    - e \vec{E}(t) \cdot \left[ \vec{r}, \rho \right] - \gamma  \left[ \vec{\mathcal{H}}\cdot \vec{s},\rho + \delta \rho(t) \right] - i\hbar \partial_t \delta \rho(t) = 0. 
\end{equation}
We shall assume that  $\rho$ describes states that are not spin-polarized 
and therefore $\left[ \vec{s}, \rho \right] = 0$ (this is the case when 
$u$ describes a spin-dependent potential which does not break time-reversal symmetry). Next we project the LQLE equation onto the basis of Bloch waves:
\begin{equation}
  \left[\delta \rho(t), h \right]^{\alpha\beta}_{kp} - \gamma \left[ \vec{\mathcal{H}}(t)\cdot \vec{s},\delta \rho(t) \right]^{\alpha\beta}_{kp} 
   - i\hbar \partial_t \delta \rho^{\alpha\beta}_{kp}(t) =  \vec{E}(t) \cdot \vec{C}^{\alpha\beta}_{kp}.
\end{equation}
We have introduced the following notation:
\begin{equation}
\left[ A, B \right]^{\alpha\beta}_{kp} = \langle k \alpha | \left[ A, B\right] | p \beta \rangle\quad;\quad \vec{C}^{\alpha\beta}_{kp} = e \left[\vec{r}, \rho \right]^{\alpha\beta}_{kp}
\end{equation}

In what follows, it will be more convenient to write the LQLE in terms of the Fourier components of the electric field and the linear
response density matrix, i.e.
\begin{align}
\left[ \delta \rho(\omega), h \right]^{\alpha\beta}_{kp} - \gamma \left[ \vec{\mathcal{H}}(\omega)\cdot \vec{s},\delta \rho(\omega) \right]^{\alpha\beta}_{kp}
-& \hbar \omega^{+} \delta \rho^{\alpha\beta}_{kp}(\omega)  \nonumber \\
&= e \vec{E}(\omega)\cdot \vec{C}^{\alpha\beta}_{kp},
\end{align}
where $\omega^{+} = \omega + 2 i \eta$ ($\eta\to 0^+$).
To proceed further, we split $\delta \rho(\omega)$ into diagonal and non-diagonal parts in the Bloch basis (but retaining the quantum coherence of the spin degree of freedom encoded in the off-diagonal components of the density matrix):
\begin{equation}
\delta \rho^{\alpha\beta}_{kp}(\omega) = \delta n^{\alpha\beta}_{k}(\omega)   + 
\delta g^{\alpha\beta}_{kp}(\omega),
\end{equation}
where
\begin{align}
\delta n^{\alpha\beta}_{k}(\omega) &= \delta \rho^{\alpha\beta}_{kk}(\omega)
\delta_{kp},\\
\delta g^{\alpha\beta}_{kp}(\omega) &= \delta \rho^{\alpha\beta}_{kp}(\omega) (1 - \delta_{kp}).
\end{align}
Note that, by construction $\delta g$ is an operator whose (momentum) diagonal elements are zero. 

The procedure by which KL derived the linearized quantum Boltzmann equation (LQBE) from the LQLE consist of integrating out the momentum off diagonal components of 
$\delta \rho(\omega)$, which are described by $\delta g(\omega)$. The first step in such elimination procedure is to derive the ``equations of motion'' for the diagonal  and off diagonal  parts from the LQLE:
\begin{align}
\left[h, \delta n(\omega) \right]^{\alpha\beta}_{kk} + \left[h, \delta g(\omega)\right]^{\alpha\beta}_{kk} -  \gamma \left[ \vec{\mathcal{H}}(\omega)\cdot \vec{s}, \delta n(\omega) \right]^{\alpha\beta}_{kk}
-\hbar \omega^{+} \delta n^{\alpha\beta}_{k}(\omega)  &= \vec{E}(\omega) \cdot \vec{C}^{\alpha\beta}_{kk},\\
 \left[h,  \delta g(\omega) \right]^{\alpha\beta}_{kp} + \left[ h, \delta n(\omega)\right]^{\alpha\beta}_{kp} -\gamma  \left[ \vec{\mathcal{H}}(\omega)\cdot \vec{s}, \delta g(\omega) \right]^{\alpha\beta}_{kp}
 - \hbar \omega^{+} \delta g^{\alpha\beta}_{kp}(\omega)  &= \vec{E}(\omega) \cdot \vec{\tilde{C}}^{\alpha\beta}_{kp}\;, (k\neq p)
\label{eq:offdiag}
\end{align}
Here we have introduced $\vec{\tilde{C}}^{\alpha\beta}_{kp} = \vec{C}^{\alpha\beta}_{kp}(1-\delta_{kp})$.

To eliminate the 
off diagonal components, we will now promote the off-diagonal part of the LQLE (i.e. Eq.~\eqref{eq:offdiag}) into an operator equation. This is done by introducing the operator,
\begin{equation}
g(\omega) = \Gamma(\omega) + \delta g(\omega),
\end{equation}
where the auxiliary operator $\Gamma^{\alpha
\beta}_{kp}(\omega) = \Gamma^{\alpha\beta}_{k}(\omega) \delta_{kp}$, that is, diagonal in
the crystal momentum indices. The operator $g(\omega)$ has the same off diagonal components of $\delta g(\omega)$ but unlike  $\delta g(\omega)$, it contains diagonal components given by $\Gamma(\omega)$. By choosing $\Gamma(\omega)$ carefully, we can turn Eq.~\eqref{eq:offdiag} into an operator equation. Physically, the operator $\Gamma(\omega)$ carries the information of scattering processes between different momentum states $p$ and $k$. In fact, as we shall see below, the collision integral of the LQBE emerges from  $\Gamma(\omega)$.

In terms of $g(\omega)$, Eq.~\eqref{eq:offdiag} reads:
\begin{equation}
\left[h, g(\omega) -\Gamma(\omega) \right]^{\alpha\beta}_{kp} + \left[ h, \delta n(\omega)\right]^{\alpha\beta}_{kp}  -\gamma  \left[ \vec{\mathcal{H}}(\omega)\cdot \vec{s},  g(\omega) - \Gamma(\omega) \right]^{\alpha\beta}_{kp} -
\hbar \omega^{+} g^{\alpha\beta}_{kp}(\omega)   =  \vec{E}(\omega)\cdot \vec{\tilde{C}}^{\alpha\beta}_{kp} \; , (k\neq p)
\end{equation}
The operator $\Gamma(\omega)$ is determined by requiring that the above equation also holds for $k = p$, which leads to:
\begin{equation}
\left[ h, g(\omega) -\Gamma(\omega) \right]^{\alpha\beta}_{kk} + \left[ h, \delta n(\omega) \right]^{\alpha\beta}_{kk} -\gamma  \left[ \vec{\mathcal{H}}(\omega)\cdot \vec{s},  g(\omega) - \Gamma(\omega) \right]^{\alpha\beta}_{kk}
- \hbar \omega \Gamma^{\alpha\beta}_{k}(\omega)   = 0 \label{eq:auxx}
\end{equation}
(recall that $\tilde{C}_{kk} = 0$ by definition). Furthermore, one can show that
the term $\gamma  \left[ \vec{\mathcal{H}}\cdot \vec{s},  g(\omega) - \Gamma(\omega) \right]^{\alpha\beta}_{kk}$ vanishes thus the condition that determines $\Gamma(\omega)$ becomes:
\begin{equation}
\left[ h, g(\omega) -\Gamma(\omega) \right]^{\alpha\beta}_{kk} + \left[ h, \delta n(\omega) \right]^{\alpha\beta}_{kk} 
- \hbar \omega^{+} \Gamma^{\alpha\beta}_{k}(\omega)   = 0 \label{eq:aux}
\end{equation}
Using this equation, we can write
the diagonal part of the LQBE, $\delta n_k(\omega)$ as follows:
\begin{equation}
  \hbar \omega^{+} \delta n^{\alpha\beta}_{k}(\omega) +
   \vec{E}(\omega)\cdot  \vec{C}^{\alpha\beta}_{kk}  
   + \gamma \left[ \vec{\mathcal{H}}(\omega)\cdot \vec{s}, \delta n_k(\omega) \right]^{\alpha\beta}
 = 
\hbar \omega^{+} \Gamma^{\alpha\beta}_{k}(\omega) .
\end{equation}
In other words, $\delta n_{k}(\omega)$ is determined by the diagonal elements of $\vec{C}$ and 
$\Gamma(\omega)$. The diagonal part of the LQLE thus resembles the form of the  Boltzmann equation, i.e. left hand side of the equation describes the drift-diffusion process while the right hand side describes the collision process.

In summary, by introducing an auxiliary matrix $\Gamma(\omega)$ which physically corresponds to the collision integral,  we have transformed the original LQBE into two operator equations given by the following:

\begin{equation} \label{eq:nk}
\hbar \omega^{+} \delta n_{k} (\omega) +
   \vec{E}(\omega)\cdot  \vec{C}_{kk}  
   + \gamma \left[ \vec{\mathcal{H}}(\omega)\cdot \vec{s}, \delta n_k(\omega) \right]
 = 
\hbar \omega^{+} \Gamma_{k}(\omega) 
\end{equation}
\begin{equation}
\left[h, g(\omega)\right] - \hbar \omega^{+} g(\omega) = \vec{E}(\omega) \cdot \vec{\tilde{C}}  + \left[ h,\Gamma(\omega) - \delta n(\omega)  \right] +\gamma  \left[ \vec{\mathcal{H}}(\omega)\cdot \vec{s},  g(\omega) - \Gamma(\omega) \right] \label{eq:geq}
\end{equation}

 In the above equations and from what follows,  the spin indices are dropped for notation simplicity. However, readers should keep in mind that $\delta n_{k}(\omega)$, $g(\omega)$, $\Gamma(\omega)$, $\vec{C}$ and $\tilde{\vec{C}}$ are matrices in spin space.

\subsubsection*{Self-consistent equation for the collision integral}

Next, we proceed with the computation of the collision integral $\hbar\omega^{+}\Gamma(\omega)$. In operator form,  $g(\omega)$ can be written as follows:
\begin{equation}
\left[h, g(\omega)\right] - \hbar \omega^{+} g(\omega)  = \Lambda(\omega),\label{eq:geqv2}
\end{equation}
where
\begin{equation}
\Lambda(\omega) = \vec{E}(\omega) \cdot \vec{\tilde{C}}  + \left[ h,\Gamma(\omega) - \delta n(\omega)  \right] +\gamma  \left[ \vec{\mathcal{H}}(\omega)\cdot \vec{s},  g(\omega) - \Gamma(\omega) \right]
\end{equation}
Note that Eq.~\eqref{eq:geqv2} must be solved self-consistently because the left-hand side depends on $g(\omega)$ as well as on $\Gamma(\omega)$, which is just a short-hand for the diagonal elements of $g(\omega)$.  To proceed further, let us write down the formal solution of $g(\omega)$ by projecting Eq.~\eqref{eq:geqv2}   onto the eigenbasis of $h$ (i.e. $h|n \rangle = \epsilon_n |n\rangle$):
\begin{equation}
\left( \epsilon_m - \epsilon_n\right) g_{mn}(\omega) - \hbar \omega^{+} g_{mn}(\omega)  = \Lambda_{mn}(\omega).
\end{equation}
Hence,
\begin{equation}
g_{mn}(\omega) = \frac{\Lambda_{mn}(\omega)}{\epsilon_m + \epsilon_n - (\hbar 
\omega - 2i \eta)},\label{eq:res1}
\end{equation}
where we have used $\omega^{+}=\omega + 2 i \eta$. Eq.~\eqref{eq:res1} can be rewritten as follows:
\begin{equation}
g_{mn}(\omega) = \frac{i}{2\pi}
\int^{+\infty}_{-\infty} \left( \frac{1}{\epsilon + \hbar\omega/2 -  \epsilon_m + i \eta}\Lambda_{mn}(\omega)\frac{1}{\epsilon-\hbar\omega/2-\epsilon_n-i\eta}\right) \, d\epsilon. \label{eq:formalsol}
\end{equation}
Then, we rewrite this equation in operator form as follows:
\begin{equation}
g(\omega) = \frac{i}{2\pi} \int d\epsilon \, 
G^{+}\left(\epsilon + \frac{\hbar\omega}{2}\right)
\Lambda(\omega) G^{-}\left(\epsilon - \frac{\hbar\omega}{2}\right),
\end{equation}
where the resolvent operators (Green functions) read ($\eta\to 0^{+}$)
\begin{equation}
G^{\pm}(\epsilon) = \frac{1}{\epsilon-h\pm i\eta}.
\end{equation}
Next, we proceed to obtain a formal solution to Eq.~\eqref{eq:formalsol}
in powers of $n_{\mathrm{imp}}$ and the strength of the external electric and magnetic fields $E(\omega)$ and $\mathcal{H}(\omega)$,
we split
\begin{equation}
g(\omega) = g_1(\omega) + g_2(\omega),
\end{equation}
and consequently, 
\begin{equation}
\Gamma(\omega) = \Gamma_1(\omega) + \Gamma_2(\omega),
\end{equation}
for the diagonal part. The operator $g_1(\omega)$  ($\Gamma_1(\omega)$) contains terms starting at
linear order in $n_{\mathrm{imp}}$ but it is zeroth
order in the external fields. $g_2(\omega)$ 
($\Gamma_2(\omega)$) contains terms starting
at $O(n^{2}_{\mathrm{imp}})$, or linear in $O(n_{\mathrm{imp}})$ but containing at least one (or higher powers) of
$|\vec{E}(\omega)|$  and/or $|\vec{\mathcal{H}}(\omega)|$. Thus, we can split
\begin{align}
\Lambda_1(\omega) &= \left[ h, \Gamma_1(\omega) - \delta n(\omega) \right] 
= \left[ u, \Gamma_1(\omega) - \delta n(\omega) \right],\\
\Lambda_2(\omega) &=  \left[u, \Gamma_2(\omega)\right]
+ \vec{E}(\omega)\cdot \vec{\tilde{C}}  + \gamma  \left[ \vec{\mathcal{H}}(\omega)\cdot \vec{s}, g_1(\omega)-\Gamma_1(\omega)\right] 
\nonumber\\
&+  \gamma  \left[ \vec{\mathcal{H}}(\omega)\cdot \vec{s}, g_2(\omega)-\Gamma_2(\omega)\right],
\end{align}
where we have used that $\left[ h_0, 
\Gamma_{1,2}(\omega)\right]  = 0$ and
$\left[h_0,\delta n(\omega) \right] = 0$. 
Hence, using Eq.~\eqref{eq:formalsol},
\begin{align}
g_1(\omega) 
= i \int \frac{d\epsilon}{2\pi} \, G^{+}\left(\epsilon+\frac{\hbar \omega}{2} \right) \left[u, \Gamma_1(\omega) -\delta n(\omega) \right]  G^{-}\left(\epsilon-\frac{\hbar \omega}{2} \right), 
\end{align}
and 
\begin{align}
g_2(\omega)  &=i \int \frac{d\epsilon}{2\pi}\, G^{+}\left(\epsilon+\frac{\hbar \omega}{2} \right) 
\Big\{ \left[u, \Gamma_2(\omega)\right]
+ \vec{E}(\omega)\cdot \vec{\tilde{C}}  + \gamma  \left[ \vec{\mathcal{H}}(\omega)\cdot \vec{s}, g_1(\omega)-\Gamma_1(\omega)\right]  
  \nonumber\\
&\qquad +\gamma  \left[ \vec{\mathcal{H}}(\omega)\cdot \vec{s}, g_2(\omega)-\Gamma_2(\omega)\right] 
\Big\}
G^{-}\left(\epsilon-\frac{\hbar \omega}{2} \right).
\end{align}

Let us consider $g_1(\omega)$, which contains the leading contributions which interest us here. For notational simplicity,  we denote
\begin{equation} \label{eq:Rshort}
R(\omega) = \delta n(\omega) - \Gamma(\omega),
\end{equation}
and    simplify $g_1(\omega)$:
\begin{align}
g_1(\omega) &= i \int \frac{d\epsilon}{2\pi} G^{+}\left(\epsilon+\frac{\hbar \omega}{2} \right) \left[ u, R(\omega)\right] G^{+}\left(\epsilon-\frac{\hbar \omega}{2} \right) \nonumber \\
&= i \int \frac{d\epsilon}{2\pi} \left[ G^{+}\left(\epsilon+\frac{\hbar \omega}{2} \right) u R(\omega) G^{-}\left(\epsilon-\frac{\hbar \omega}{2} \right)  - G^{+}\left(\epsilon+\frac{\hbar \omega}{2} \right)  R(\omega) u G^{-}\left(\epsilon-\frac{\hbar \omega}{2} \right) \right] \nonumber\\
&= i \int \frac{d\epsilon}{2\pi} \left[ G^{+}_0\left(\epsilon+\frac{\hbar \omega}{2} \right) T^{+}\left(\epsilon + \frac{\hbar\omega}{2}\right) R(\omega) G^{-}\left(\epsilon-\frac{\hbar \omega}{2} \right) \right.\nonumber \\
&\qquad \left. - G^{+}\left(\epsilon+\frac{\hbar \omega}{2} \right)  R(\omega) T^{-}\left(\epsilon - \frac{\hbar \omega}{2}\right) G^{-}_0\left(\epsilon-\frac{\hbar \omega}{2} \right) \right].
\end{align}
Here the relationships between $T$-matrix and Green function, $G^{+}u=G_{0}T^{+}$ and  $uG^{-}=T^{-}G_{0}$ are used. 
Next, using the Dyson  equation $G^{\pm}(\epsilon) = G^{\pm}_0(\epsilon) + G^{\pm}_0(\epsilon) T^{\pm}(\epsilon) G^{\pm}_0(\epsilon)$, we split the above integral in two terms $g_1(\omega) = I^{(1)}(\omega) + I^{(2)}(\omega)$, where
\begin{align}
I^{(1)}(\omega) &= i \int \frac{d\epsilon}{2\pi} \left[ G^{+}_0\left(\epsilon+\frac{\hbar \omega}{2} \right) T^{+}\left(\epsilon + \frac{\hbar\omega}{2}\right) R(\omega) G^{-}_0\left(\epsilon-\frac{\hbar \omega}{2} \right) \right. \nonumber  \\
&\qquad \left. - G^{+}_0\left(\epsilon+\frac{\hbar \omega}{2} \right)  R(\omega) T^{-}\left(\epsilon - \frac{\hbar \omega}{2}\right) G^{-}_0\left(\epsilon-\frac{\hbar \omega}{2} \right) \right]. \\
I^{(2)}(\omega) &= i \int \frac{d\epsilon}{2\pi} \left[ G^{+}_0\left(\epsilon+\frac{\hbar \omega}{2} \right) T^{+}\left(\epsilon + \frac{\hbar\omega}{2}\right) R(\omega)  G^{-}_{0}\left( \epsilon - \frac{\hbar\omega}{2} \right)  T^{-}\left(\epsilon - \frac{\hbar\omega}{2}\right) G^{-}_0\left(\epsilon-\frac{\hbar \omega}{2} \right) \right. \nonumber \\
&\qquad \left. - G^{+}_0\left(\epsilon+\frac{\hbar \omega}{2} \right) 
T^{+}\left(\epsilon + \frac{\hbar \omega}{2}\right)  G^{+}_{0}\left( \epsilon + \frac{\hbar\omega}{2} \right)  R(\omega)  T^{-}\left(\epsilon - \frac{\hbar \omega}{2}\right)  G^{-}_0\left(\epsilon-\frac{\hbar \omega}{2} \right) \right]. \nonumber\\
\end{align}
Note that by construction, $R(\omega)$  is diagonal in momentum indices. If we take the diagonal (in crystal momentum indices) elements :
\begin{align}
\left(I^{(1)}(\omega)\right)_{kk}=  
 i \int \frac{d\epsilon}{2\pi} \left( G^{+}_{0k}\left(\epsilon+\frac{\hbar \omega}{2} \right) G^{-}_{0k}\left(\epsilon-\frac{\hbar \omega}{2} \right) \right)
\left[ T^{+}\left(\epsilon + \frac{\hbar\omega}{2}\right) R(\omega) - R(\omega) T^{-}\left(\epsilon - \frac{\hbar \omega}{2}\right)  \right]_{kk}. 
\end{align}
Note that the big parenthesis containing $G_{0k}^{+}$ and $G_{0k}^{-}$ is a number whereas the terms inside the square bracket is a matrix in spin space. To evaluate this integral, we use 
\begin{align}
\Pi_{k}(\epsilon,\hbar \omega) &= (\hbar \omega + 2 i \eta) G^{+}_{0k}\left(\epsilon+\frac{\hbar \omega}{2} \right) 
G^{-}_{0k}\left(\epsilon-\frac{\hbar \omega}{2} \right) \nonumber \\
 &= (\hbar \omega + 2 i \eta) \left(\frac{1}{\epsilon+\hbar\omega/2 - \epsilon_k + i\eta} \right) \left( \frac{1}{\epsilon-\hbar\omega/2 - \epsilon_k - i\eta}\right) \nonumber\\ 
&=  \left[ \frac{1}{\epsilon+\hbar\omega/2 - \epsilon_k + i\eta} - \frac{1}{\epsilon-\hbar\omega/2 - \epsilon_k - i\eta}
\right] \nonumber \\
&= G^{-}_{0k}\left(\epsilon-\frac{\hbar\omega}{2}\right) - G^{+}_{0k}\left(\epsilon+\frac{\hbar\omega}{2}\right) .
\end{align}
Hence,
\begin{equation}
\hbar \omega^+ \left(I^{(1)}(\omega)\right)_{kk}= i  \int \frac{d\epsilon}{2\pi} 
 G^{-}_{0k}\left(\epsilon-\frac{\hbar\omega}{2}\right)  \left[T^{+}\left(\epsilon+
 \frac{\hbar\omega}{2} \right) R(\omega) \right]_{kk}
  + G^{+}_{0k} \left( \epsilon + \frac{\hbar\omega}{2}\right)
\left[ R(\omega)  T^{-}\left(\epsilon - \frac{\hbar\omega}{2}\right) \right]_{kk} . \label{eq:i1int}
\end{equation}
Note that a term of the form 
\begin{equation*}
\int \frac{d\epsilon}{2\pi} G^{+}_{0k}\left(\epsilon+\frac{\hbar\omega}{2}\right)  \left[T^{+} \left(\epsilon+\frac{\hbar\omega}{2} \right) R(\omega) \right]_{kk}
\end{equation*}
vanishes because the poles all lie on the lower half-plane. Closing the contour on the upper-half side of the complex plane therefore picks up a vanishing contribution. Similar remarks apply to the term of the form:
\begin{equation*}
\int \frac{d\epsilon}{2\pi} G^{-}_{0k}\left(\epsilon-\frac{\hbar\omega}{2}\right) \left[ R(\omega) T^{-} \left(\epsilon-\frac{\hbar\omega}{2} \right) \right]_{kk}.
\end{equation*}
Thus, we can proceed to evaluate the integral over $\epsilon$ in Eq.~\eqref{eq:i1int}. For the first term, we choose to close the contour in the 
upper half-plane where the pole of $G^{-}_{0k}\left(\epsilon - \frac{\hbar\omega}{2}\right)$ lies at $\epsilon = \epsilon_k + \frac{\hbar\omega}{2} + i\eta$. For the second term, we choose to close the contour in the lower
half-plane, where the pole of $G^{+}_{0k}\left(\epsilon + \frac{\hbar\omega}{2}\right)$ lies at $\epsilon = \epsilon_k - \frac{\hbar\omega}{2} - i\eta$. 
Therefore, Eq.~\eqref{eq:i1int} becomes
\begin{align}
\hbar \omega^+ (I^{(1)}(\omega))_{kk}
=& \left[ R (\omega) T^{+}(\epsilon_k + \hbar\omega) 
- T^{-} \left(\epsilon_k - \hbar \omega\right) R (\omega) \right]_{kk}.
\end{align}
We would like to remind the reader that both the $T$ matrix and the operator $R$ are matrices in spin space and their ordering is important. In terms of their components (where
summation over repeated indices is implied):
\begin{equation}
\hbar \omega^+ (I^{(1)}(\omega))_{kk}^{\alpha \beta}= R^{\alpha\gamma}_{k}(\omega) (T^{+})^{\gamma\beta}_{kk}(\epsilon_k + \hbar\omega) 
- (T^{-})^{\alpha\gamma}_{kk}\left(\epsilon_k - \hbar \omega\right) R^{\gamma\beta}_{k}(\omega).
\end{equation}
Recall that the operator $R$ is diagonal in $k$ by construction; repeated Greek indices are summed.

Let us now take up the second integral $I^{(2)}(\omega)$. Again,
we focus on the diagonal elements of momentum which are needed to determine the
equation for the diagonal part of the density matrix, $\delta n(\omega)$:

\begin{align}
\hbar\omega^{+} \left(I^{(2)}(\omega) \right)_{kk} 
=   i \sum_{p} \int \frac{d\epsilon}{2\pi}\,  \Pi_k(\epsilon,\hbar \omega) \Pi_p(\epsilon,\hbar\omega) \left[
 T^{+}_{kp}\left(\epsilon + \frac{\hbar\omega}{2} \right)
 R_{p}(\omega) T^{-}_{pk}\left(\epsilon - \frac{\hbar\omega}{2}\right) \right]
\end{align}

%
In the last line, the terms in the square bracket is a matrix in spin space. Next, let us study the function
\begin{equation}
\Pi_k(\epsilon,\hbar\omega)  =  G^{-}_{0k} \left(\epsilon - \frac{\hbar \omega}{2} \right) - G^{+}_{0k} \left(\epsilon + \frac{\hbar \omega}{2} \right).
\end{equation}
Typically the values of $k$ and $\hbar\omega$ for which we are interested 
in this function are such that $k \approx k_F$ (i.e. $\epsilon_k \approx \mu$) and $\hbar \omega \ll \mu$. Thus it is possible to expand this function in powers of $\hbar \omega$. The leading order term is obtained by setting $\hbar \omega = 0$:
\begin{equation}
\Pi_k(\epsilon,\hbar\omega = 0) = \left[ G^{-}_{0k}  \left(\epsilon \right) - G^{+}_{0k} \left(\epsilon \right)\right] = 2 i \pi \delta(\epsilon-\epsilon_k).
\end{equation}
We shall see below that $\hbar \omega^{+} I^{(2)}_{kk}(\omega)$ is of linear order in $n_{\mathrm{imp}}$ and we shall be neglecting terms of order $O\left[n_{\mathrm imp} (\hbar\omega/\mu)\right]$ assuming that $\hbar\omega \ll \mu$. Thus, we approximate
\begin{equation}
\Pi_k(\epsilon,\hbar\omega) \approx \Pi_k(\epsilon,\hbar\omega =0) =
2i \pi \delta(\epsilon-\epsilon_k).
\end{equation}
Hence, 
\begin{equation}
\hbar\omega^{+} \left(I^{(2)}(\omega)\right)_{kk} \approx 
- 2\pi i \sum_{p} T^{+}_{kp}\left(\epsilon_p \right) R_{p}(\omega) T^{-}_{pk}\left(\epsilon_p\right) \delta(\epsilon_p -\epsilon_k).
\end{equation}
This  is the quantum analogue of the scattered-in term of the collision integral in the semiclassical Boltzmann equation. 

Therefore,  collecting the two results  $\hbar (\omega^{+} I^{(1)}(\omega))_{kk}$ and  $\hbar\omega^{+} (I^{(2)}(\omega))_{kk}$ together, we arrive at the following equation:
\begin{align}
 \hbar \omega^{+} \Gamma_{1k}(\omega) &= \left[ \delta n_{k}(\omega) -\Gamma_{1k}(\omega) \right] T^{+}_{kk}(\epsilon_k) 
- T^{-}_{kk}\left(\epsilon_k\right) \left[ \delta n_{k}(\omega) -\Gamma_{1k}(\omega) \right] \nonumber \\
&- 2\pi i \sum_{p}  T^{+}_{kp}\left(\epsilon_p \right)
\left[ \delta n_{p}(\omega) -\Gamma_{1p}(\omega) \right]  T^{-}_{pk}\left(\epsilon_p\right) \delta(\epsilon_p -\epsilon_k). \label{eq:gamma1}
\end{align}
Note we have used the definition $R_{k}=\delta n_{k}- \Gamma_{1k}$ defined in Eq~\eqref{eq:Rshort}.  This is the self-consistent equation of the collision integral $\Gamma_{1k}$. An exact solution of the above equation appears to be very difficult, as it requires the knowledge of the multiple impurity $T$-matrix. To get around this obstacle, following KL, we solve this equation in powers of the impurity density $n_{\mathrm{imp}}$. 

\subsubsection*{Virial expansion}

We first notice that in powers of  $n_{\mathrm{imp}}$, the $T$-matrix admits the following virial expansion
(see Ref.~\onlinecite{KohnLuttingerBTE}). The volume of the system
in this section is taken to be unity:
\begin{equation}
T^{\pm}(\epsilon) = \sum_{i} T^{\pm}_{i}(\epsilon) + \frac{1}{2} \sum_{i\neq j}
\left[ T^{\pm}_{ij}(\epsilon) - T^{\pm}_{i}(\epsilon) - T^{\pm}_{j}(\epsilon) \right] + \cdots
\end{equation}
where the terms in the above series are arranged in increasing
order in $n_{\mathrm{imp}}$; $T_{i}$  being the single impurity $T$-matrix for the impurity located at $r = r_i$.  Similarly, $T_{ij}$ is the $T$-matrix of two impurities located at $r_{i}$ and $r_{j}$. The relevant matrix elements in Eq. \eqref{eq:gamma1} are:

\begin{equation}
T^{\pm}_{kk}=  \sum_{i} \langle k | T^{\pm}_{i}  | k \rangle  + \cdots
= n_{\mathrm{imp}} \mathcal{T}^{\pm}_{kk}(\epsilon) + O(n_{\mathrm{imp}}^2),
\end{equation}

\begin{align}
T^{+}_{kp} \Gamma_{1p} T^{-}_{pk}=&
\left( \sum_{i} \left( T^{+}_{i}  \right)_{kp} + \cdots  \right) \Gamma_{1p}
\left( \sum_{j} \left(T^{-}_{j} \right)_{pk} + \cdots  \right) \nonumber \\
 =&  \sum_{i}     \left( T^{+}_{i}  \right)_{kp} \Gamma_{1p}\left(T^{-}_{i} \right)_{pk} + \sum_{i\neq j } \left( T^{+}_{i}  \right)_{kp} \Gamma_{1p}\left(T^{-}_{j} \right)_{pk} + \cdots \nonumber \\
 =&  n_{\mathrm{imp}} \, \mathcal{T}^{+}_{kp} \Gamma_{1p}  \mathcal{T}^{-}_{pk}  + O(n_\mathrm{imp}^2).
\end{align}

Here $\mathcal{T}^{\pm}_{kp}$ is the matrix element of a $T$-matrix for a single impurity located at the origin $r_i =0$.  In the above derivation, we used the translational property of a single impurity,
\begin{equation}
 \left(T^{\pm}_{i} \right)_{kp} = \langle k | T^{\pm}(r-r_{i})|p\rangle =\langle k |e^{-i \hat{q} r_{i}} T^{\pm}(r) e^{i \hat{q} r_{i}}|p\rangle  = e^{i (p-k)r_i}  \mathcal{T}_{kp}^{\pm} .
\end{equation}

To proceed further, we see from Eq.~\eqref{eq:gamma1} that the series for $\Gamma_1(\omega)$ starts at linear order in $n_{\mathrm{imp}}$,  hence we expand:
\begin{equation}
\Gamma_1(\omega) = A_{1}(\omega) + A_{2}(\omega) + \cdots
\end{equation}
where $A_{m}(\omega)$ is $O(n^m_{\mathrm{imp}})$. Similarly, we can
expand:
\begin{equation}
\Gamma_2(\omega) = \left[ B_{2}(\omega) + B_{3}(\omega) + \cdots \right] 
+ \left[ D_{1}(\omega) + D_{2}(\omega) + \cdots \right]
\end{equation}
where the first set of terms in brackets $B_{m}(\omega)$ ($m \ge 2$) are 
$O(n^m_{\mathrm{imp}})$ ; the second set of terms in the brackets $D_{m}$ is a force term which is  $O(|\vec{F}(\omega)| n^m_{\mathrm{imp}})$, with $\vec{F}(\omega) = \vec{E}(\omega), \vec{\mathcal{H}}(\omega)$ and $m\ge 1$.
 Therefore, to leading order in
$n_{\mathrm{imp}}$ and zeroth order in $n_{\mathrm{imp}}|\vec{F}|$,  $\Gamma_{2}(\omega)=0$ and the collision integral Eq. \eqref{eq:gamma1} becomes:
\begin{align}
\hbar \omega^{+} \Gamma_{1k}(\omega) =&  n_{\mathrm{imp}}\, \left( \delta n_{k}(\omega) \mathcal{T}^{+}_{kk}(\epsilon_k) 
- \mathcal{T}^{-}_{kk}\left(\epsilon_k\right)  \delta n_{k}(\omega) \right) \nonumber \\
&- 2\pi n_{\mathrm{imp}}\, i \sum_{p}  \mathcal{T}^{+}_{kp}\left(\epsilon_p \right)
 \delta n_{p}(\omega)\mathcal{T}^{-}_{pk}\left(\epsilon_p\right) \delta(\epsilon_p -\epsilon_k).
\end{align}
Lastly, we wish to express the collision integral of the QBE as follow: 
\begin{align}
 &\quad \delta n_{k}(\omega) \mathcal{T}^{+}_{kk}(\epsilon_k) 
- \mathcal{T}^{-}_{kk}\left(\epsilon_k\right)  \delta n_{k}(\omega)  \nonumber \\
 &=   \delta n_{k}(\omega) \Sigma^{R}_{k}(\epsilon_k) 
- \Sigma^{R}_{k}(\epsilon_k)  \delta n_{k}(\omega) 
+  i \left( \delta n_{k}(\omega) \Sigma^{I}_{k}(\epsilon_k) 
 + \Sigma^{I}_{k}(\epsilon_k)  \delta n_{k}(\omega)  \right) \nonumber \\
\end{align}
where $\Sigma^R_{k}(\epsilon) = \left[\mathcal{T}^{+}_{kk}(\epsilon) + \mathcal{T}^{-}_{kk}(\epsilon)\right]/2$ and
$\Sigma^I_{k}(\epsilon) = \left[\mathcal{T}^{+}_{kk}(\epsilon) - \mathcal{T}^{-}_{kk}(\epsilon)\right]/(2i)$ are the hermitian and anti-hermitian part of the single impurity $T$-matrix. From optical theorem, the anti-hermitian part of the $T$-matrix is related to the scattering cross section, 
\begin{equation}
\Sigma^I_{k}(\epsilon_k) = \pi \sum_{p} \mathcal{T}^{+}_{kp}(\epsilon_k)  \mathcal{T}^{-}_{pk}(\epsilon_k) \delta(\epsilon_{k}-\epsilon_{p}). 
\end{equation}
Therefore, the collision integral can be written into the more familiar form,
\begin{align}
\label{eq:coll-int2}
\hbar \omega^{+} \Gamma_{1k}(\omega)=&  [\delta n_{k}(\omega) ,\Sigma^R_{k}(\epsilon_{k}) ] - \frac{2\pi i n_{\mathrm{imp}}}{\hbar} \sum_{\vec{p}} \delta(\epsilon_{k}-\epsilon_{p})  \times \nonumber \\
 & \left( \mathcal{T}^{+}_{kp} (\epsilon_p) \delta n_{p}(\omega)  \mathcal{T}^{-}_{pk}(\epsilon_p)-
\frac{ \mathcal{T}^{+}_{kp} (\epsilon_k) \mathcal{T}^{-}_{pk}(\epsilon_k)  \delta n_{k}(\omega) + 	\delta n_{k}(\omega)  \mathcal{T}^{+}_{kp}(\epsilon_k)  \mathcal{T}^{-}_{pk}(\epsilon_k) }{2} \right).
\end{align}

The above expression is the collision integral of the QBE apart from a factor of $-i$. Before we substitute this equation back to the diagonal part of LQBE ( Eq. ~\eqref{eq:nk}) and arrive at the linearized spin-coherent QBE, the left hand side of Eq.  \eqref{eq:nk} must  (also) expand to leading order in $n_{imp}$ and zeroth order in $n_{imp}F(\omega)$ (for $F(\omega)=E(\omega),\mathcal{H}(\omega)$):
\begin{align}
&\quad \hbar\omega^{+}  \delta n_{k}(\omega) 
 + \gamma \left[ \boldsymbol{\mathcal{H}}(\omega)\cdot \vec{s}, \delta n_k(\omega) \right]
+  \vec{E}(\omega) \cdot (\vec{C})_{kk}  \nonumber \\
&= \quad \hbar\omega^{+}  \delta n_{k}(\omega) 
 + \gamma \left[ \boldsymbol{\mathcal{H}}(\omega)\cdot \vec{s}, \delta n_k(\omega) \right]
+  \vec{E}(\omega) \cdot (\vec{C}_{0})_{kk} .
\label{eq:lqbe}
\end{align}
Here, we have expanded the force term in powers of $n_{\mathrm{imp}}$,
\begin{equation}
\vec{C} = e \left[\vec{r}, \rho \right] = \vec{C}_0  + \vec{C}_1 + \cdots 
\end{equation}
where the zeroth order term is of the form,
\begin{equation}
\vec{C}_0 = e \left[ \vec{r}, \rho_0\right].
\end{equation}
Here $\rho_{0}=(e^{(h_{0}-\mu)/k_B T} +1)^{-1}$ is the Fermi-Dirac distribution function in the \emph{absence} of impurities. The matrix components of $\vec{C}_0$ is given by
\begin{equation}
 (\vec{C}_{0})_{kk} = e \left[ \vec{r}, \rho_0 \right]_{kk} = 
e  \:\langle k | \left[ \vec{r}, \rho_0 \right] | k  \rangle = ie\: \vec{\nabla}_k n^0(\epsilon_k) \times \mathbb{I}, 
\end{equation}
where we have used that $[\vec{r},F(\vec{k})] = i\vec{\nabla}_k F(\vec{k})$ and $\mathbb{I}$ is the unit matrix in spin space.
This result follows e.g. from the commutation relation $[r_i, k_j] = i \delta_{ij}$. Hence, for any analytic function of the (crystal) momentum $\vec{k}$ (in units of $\hbar$), it is possible to expand in Taylor series around $\vec{k} = \vec{0}$  and obtain $[r_i,f(\vec{k})] = i\partial_{k_i} f(\vec{k})$. Notice that for graphene, there is no Berry connection contribution away from the Dirac point.
 Collecting all terms, we finally arrive at the (linearized) QBE: 
\begin{equation}
-i \omega^{+}  \delta n^{\alpha\beta}_{k}(\omega) 
 - \frac{i}{\hbar} \gamma \left[ \boldsymbol{\mathcal{H}}(\omega)\cdot \vec{s}, \delta n_k(\omega) \right]^{\alpha\beta} 
-  e \vec{E}(\omega)\cdot  \frac{\vec{\nabla}_{k} n^0(\epsilon_k)}{\hbar} \delta^{\alpha\beta} =
\mathcal{I}[\delta n_k(\omega)]  
\end{equation}
or in real time:
\begin{equation} 
\partial_t \delta n^{\alpha\beta}_{k}(t) 
 - \frac{i}{\hbar} \gamma \left[ \boldsymbol{\mathcal{H}}(t)\cdot \vec{s}, \delta n_k(t) \right]^{\alpha\beta} 
-  e \vec{E}(t)\cdot  \frac{\vec{\nabla}_{k}  n^0(\epsilon_k)}{\hbar} \delta^{\alpha\beta} =
\mathcal{I}[\delta n_k(t)]
\end{equation}
The collision integral $\mathcal{I}[\delta n_k(s)]$ to leading order in $n_{\mathrm{imp}}$ takes
the form ($s \equiv \omega, t$):
\begin{align} \label{eq:result_collint}
\mathcal{I}[\delta n_k(s)] =&  \frac{i}{\hbar}\left[  \Sigma_{k}^{R}(\epsilon_k), \delta n_k(s)  \right]^{\alpha\beta} \nonumber \\
+&  \, \delta n^{\alpha\gamma}_k(s) \left[ \frac{\pi}{\hbar}  \sum_{p} \left( \mathcal{T}^{+}(\epsilon_p)\right)^{\gamma \theta}_{kp} (\mathcal{T}^{-}(\epsilon_p))^{\theta \beta}_{pk}   \delta(\epsilon_{p}-\epsilon_{k}) \right] \nonumber\\
 +&   \left[ \frac{\pi}{\hbar}   \sum_{p} (\mathcal{T}^{+} (\epsilon_p))^{\alpha\gamma}_{kp} (\mathcal{T}^{-} (\epsilon_p))^{\gamma \theta}_{pk}  
 \delta(\epsilon_{p}-\epsilon_{k})   \right] \delta n^{\theta \beta}_{k}(s)
\nonumber\\
-& \frac{2\pi}{\hbar} \sum_{p} (\mathcal{T}^{+}(\epsilon_p))^{\alpha\gamma}_{kp}  \delta n^{\gamma\theta}_{k}(s)  (\mathcal{T}^{-}(\epsilon_p))^{\theta \beta}_{pk}  \delta(\epsilon_p -\epsilon_k). 
\end{align}
The self-energy correction term is given by the hermitian part of the $T$-matrix,
\begin{equation}
\Sigma^R_k(\epsilon) = \frac{\mathcal{T}^{+}_{kk}(\epsilon) + \mathcal{T}^{-}_{kk}(\epsilon)}{2} 
\end{equation}
At the end of this section, the spin labels (greek letters) are reinserted to remind the readers that the disorder is spin dependent. 
\end{widetext}

\bibliography{reference2}

\end{document}